 \documentclass[final,5p,times,twocolumn,authoryear]{elsarticle}

\usepackage{amssymb}
\usepackage{amsmath}
\usepackage{lipsum}
\usepackage{float}
\usepackage{subcaption}
\usepackage{comment}
\usepackage{newtxtext,newtxmath}
\usepackage{hyperref}
\usepackage{xcolor}

\journal{High Energy Astrophysics}

\begin{document}

\begin{frontmatter}



\title{On the statistical assumption on the distance moduli of Supernovae Ia and its impact on the determination of cosmological parameters}


\author[first,second,third,fourth]{M.G. Dainotti \corref{cor}}
\author[fifth,sixth]{G. Bargiacchi}
\author[seventh,eighth]{M. Bogdan}
\author[fifth,sixth,ninth]{S. Capozziello}
\author[tenth,eleventh,twelfth]{S. Nagataki}

\cortext[cor]{The first and second author have contributed equally to the paper. Corresponding author,  maria.dainotti@nao.ac.jp.}

\affiliation[first]{organization={National Astronomical Observatory of Japan},
            addressline={2 Chome-21-1 Osawa, Mitaka}, 
            city={Tokyo},
            postcode={181-8588}, 
            country={Japan}}
\affiliation[second]{organization={The Graduate University for Advanced Studies, SOKENDAI},
            addressline={Shonankokusaimura, Hayama, Miura District}, 
            city={Kanagawa},
            postcode={240-0193}, 
            country={Japan}}
\affiliation[third]{organization={Space Science Institute},
            addressline={4765 Walnut St, Suite B}, 
            city={Boulder},
            postcode={80301}, 
            state={Colorado},
            country={USA}}
\affiliation[fourth]{organization={Department of Physics and Astrophysics, University of Las Vegas},
            postcode={89154}, 
            country={USA}}
\affiliation[fifth]{organization={Scuola Superiore Meridionale},
            addressline={Largo S. Marcellino 10}, 
            city={Naples},
            postcode={80138}, 
            country={Italy}}
\affiliation[sixth]{organization={Istituto Nazionale di Fisica Nucleare (INFN), Sez. di Napoli},
            addressline={Complesso Univ. Monte S. Angelo, Via Cinthia 9}, 
            city={Naples},
            postcode={80126}, 
            country={Italy}}
\affiliation[seventh]{organization={University of Wroclaw},
            addressline={plac Grunwaldzki 2/4}, 
            city={Wroclaw},
            postcode={50-384}, 
            country={Poland}}
\affiliation[eighth]{organization={Department of Statistics, Lund University},
            addressline={Box 117, SE-221 00}, 
            city={Lund},
            country={Sweden}}
\affiliation[ninth]{organization={Dipartimento di Fisica "E. Pancini", Università degli Studi di Napoli Federico II},
            addressline={Complesso Univ. Monte S. Angelo, Via Cinthia 9}, 
            city={Naples},
            postcode={80126}, 
            country={Italy}}
\affiliation[tenth]{organization={Interdisciplinary Theoretical \& Mathematical Science Program, RIKEN (iTHEMS)},
            addressline={2-1 Hirosawa, Wako, Saitama}, 
            postcode={351-0198}, 
            country={Japan}}
\affiliation[eleventh]{organization={RIKEN Cluster for Pioneering Research, Astrophysical Big Bang Laboratory (ABBL)},
            addressline={2-1 Hirosawa, Wako, Saitama}, 
            postcode={351-0198}, 
            country={Japan}}
\affiliation[twelfth]{organization={Astrophysical Big Bang Group (ABBG), Okinawa Institute of Science and Technology Graduate University (OIST)},
            addressline={1919-1 Tancha, Okinawa}, 
            postcode={904-0495}, 
            country={Japan}}

\begin{abstract}
Type Ia Supernovae (SNe Ia) are considered the most reliable \textit{standard candles} and they have played an invaluable role in cosmology since the discovery of the Universe's accelerated expansion. During the last decades, the SNe Ia samples have been improved in number, redshift coverage, calibration methodology, and systematics treatment. These efforts led to the most recent \textit{``Pantheon"} (2018) and \textit{``Pantheon +"} (2022) releases, which enable to constrain cosmological parameters more precisely than previous samples. In this era of precision cosmology, the community strives to find new ways to reduce uncertainties on cosmological parameters. 
To this end, we start our investigation even from the likelihood assumption of Gaussianity, implicitly used in this domain. Indeed, the usual practise involves constraining parameters through a Gaussian distance moduli likelihood. This method relies on the implicit assumption that the difference between the distance moduli measured and the ones expected from the cosmological model is Gaussianly distributed. In this work, we test this hypothesis for both the \textit{Pantheon} and \textit{Pantheon +} releases. We find that in both cases this requirement is not fulfilled and the actual underlying distributions are a logistic and a Student's t distribution for the \textit{Pantheon} and \textit{Pantheon +} data, respectively. When we apply these new likelihoods fitting a flat $\Lambda$CDM model, we significantly reduce the uncertainties on the matter density $\Omega_M$ and the Hubble constant $H_0$ of $\sim 40 \%$. As a result, the Hubble tension is increased at $> 5 \sigma$ level. This boosts the SNe Ia power in constraining cosmological parameters, thus representing a huge step forward to shed light on the current debated tensions in cosmology.
\end{abstract}




\end{frontmatter}




\section{Introduction}
\label{introduction}

The most accredited Universe parametrization is the $\Lambda$CDM model \citep{peebles1984}, which includes a cold dark matter (CDM) component and a cosmological constant $\Lambda$ \citep{2001LRR.....4....1C}, as suggested by the Universe accelerated expansion \citep{riess1998,perlmutter1999}. Predictions from this scenario are compatible with the cosmic microwave background \citep[CMB; ][]{planck2018}, the baryon acoustic oscillations \citep[][]{eboss2021}, and type Ia Supernovae (SNe Ia). Nevertheless, the $\Lambda$CDM model suffers from theoretical and observational shortcomings: the cosmological constant problem \citep{1989RvMP...61....1W}, the fine-tuning, the origin and properties of dark energy, and the Hubble constant ($H_0$) tension, which is the discrepancy between the $H_0$ value obtained from the Planck data of the CMB within a flat $\Lambda$CDM model ($H_0 = 67.4 \pm 0.5  \, \mathrm{km} \, \mathrm{s}^{-1} \, \mathrm{Mpc}^{-1}$, \citealt{planck2018}), and the local $H_0$ from SNe Ia and Cepheids ($H_0 = 73.04 \pm 1.04  \, \mathrm{km} \, \mathrm{s}^{-1} \, \mathrm{Mpc}^{-1}$, \citealt{2022ApJ...934L...7R}). This discrepancy varies from 4.4 to 6 $\sigma$ when different samples are considered \citep{2019ApJ...876...85R,2020PhRvR...2a3028C,2020MNRAS.498.1420W}.
Time-delay and strong lensing from quasars \citep{2019ApJ...886L..23L} provide a value close to SNe Ia, while cosmic chronometers favour the $H_0$ of the CMB \citep{2018JCAP...04..051G}. Instead, quasars \citep{rl19,2020A&A...642A.150L,biasfreeQSO2022}, the Tip of the Red-Giant Branch \citep{2021ApJ...919...16F}, and Gamma-ray bursts \citep{2009MNRAS.400..775C,cardone10,Dainotti2013a, Postnikov14, 2021ApJ...912..150D,2022Galax..10...24D,2022MNRAS.tmp.2639D,2022PASJDainotti}, hint at a $H_0$ value halfway between these two.
So, the $H_0$ tension remains an open question. One way to better highlight the tension is to reduce $H_0$ uncertainty.
As a first step towards this goal, we have questioned the Gaussianity assumption on the likelihood traditionally used for SNe Ia.
Indeed, the common practise of constraining cosmological parameters with a Gaussian distance moduli likelihood, $\cal L$$_{Gauss}$, relies on the implicit assumption that the difference between the distance moduli, $\mu$, measured and expected from a cosmological model, normalized to the uncertainty, is normally distributed.
Nonetheless, the weak lensing effect should cause a skew in the supernova magnitude distribution, which increases with redshift, thus leading to a non-Gaussianity of the supernova distribution \citep{2005ApJ...631..678H}. As a consequence, the amount of skewness can be also used to measure the amount of lensing, and test whether it is consistent with the measured clustering, as shown in \citet{2010PhRvL.105l1302A,2010MNRAS.405..535J,2014ApJ...780...24S,2020MNRAS.496.4051M}.
In this scenario of non-Gaussianity, the appropriate choice for the cosmological likelihood is crucial to obtain statistically reliable results and better constrain cosmological parameters, reducing their uncertainties. Hence, we here investigate this assumption, for the \textit{Pantheon} \citep{scolnic2018} and \textit{Pantheon +} \citep{pantheon+} samples, determining the best-fit likelihoods which we then employ to properly perform a cosmological study and constrain cosmological parameters with each of the SNe Ia data.

The manuscript is composed as follows. 
Section \ref{data} introduces the SNe Ia samples and the use of distance moduli in SNe Ia cosmology. Then, Section \ref{normalitytests} details the Gaussianity assumption along with the Gaussianity tests performed. Section \ref{fit} describes the fitting procedure. In Section \ref{results}, we describe the results both for the best-fit likelihoods and their cosmological application. In Section \ref{conclusions} we draw our conclusions. In \ref{appendix} we test the reliability of our results through mock samples.

\section{Methodology}
\subsection{Data and the  distance moduli}
\label{data}
We use the ``\textit{Pantheon}" \citep{scolnic2018} and the ``\textit{Pantheon +}" \citep{pantheon+} samples considering the covariance matrix with statistical and systematic uncertainties.
The first includes 1048 sources in the redshift range $z=0.01 - 2.26$, from CfA1-4, \textit{Carnegie Supernova Project}, \textit{Pan-STARRS1}, \textit{Sloan Digital Sky Survey}, \textit{Supernova Legacy Survey}, and \textit{Hubble Space Telescope}, 
the latter counts 1701 SNe Ia from 18 different surveys in the range $z=0.001 - 2.26$.
The \textit{Pantheon +} sample has an increased redshift span, and number of sources at low redshift, and an enriched treatment of systematic uncertainties. 
The ``\textit{Pantheon+}" contains 753 more SNe Ia compared to the \textit{Pantheon} which are not present in the \textit{Pantheon}, while the Pantheon sample has 182 SNe Ia which are not included in the ``\textit{Pantheon+}".
These improvements enable better constraints on cosmological parameters \citep[][]{2022ApJ...938..110B}.
We use both samples to investigate how and to what extent our analysis is impacted by the
above-mentioned changes in the SNe Ia sample.


To apply SNe Ia as cosmological tools, we compute the difference between the observed and the theoretical distance moduli, $\mu$.
 The observed $\mu$ is
\begin{equation}
\label{muobs}
\mu_{obs} = m_B - M + \alpha x_1 - \beta c + \Delta_M + \Delta_B
\end{equation}
where $m_B$ is the B-band overall amplitude, $x_1$ the stretch parameter, $c$ the color, $\alpha$ and $\beta$ the coefﬁcients of the relation of luminosity with $x_1$ and $c$, respectively, $M$ the fiducial B-band absolute magnitude of a SN with $x_1 = 0$ and $c=0$, $\Delta_M$ and $\Delta_B$ the corrections on the distance accounting for host-galaxy mass and biases, predicted through simulations,
respectively \citep{scolnic2018,pantheon+}.  SNe Ia alone cannot determine $H_0$, as it is degenerate with $M$, whereas $M$ depends on biases, contributions to statistical and systematic errors, etc. 
\citep{scolnic2018}.
Nevertheless, $H_0$ can be derived by fixing $M$.
Usually, Cepheids are used to determine $M$ \citep{2022ApJ...938..110B,2022ApJ...934L...7R}. \citet{scolnic2018} fix $M=-19.35$, as can be derived by the inversion of Eq. \eqref{muobs},
implying $H_0 = 70 \, \mathrm{km} \, \mathrm{s}^{-1} \, \mathrm{Mpc}^{-1}$ \citep[][]{2021ApJ...912..150D,2022Galax..10...24D}. \citet{2022ApJ...934L...7R} obtain $M = -19.253 \pm 0.027$ and $H_0 = 73.04 \pm 1.04  \, \mathrm{km} \, \mathrm{s}^{-1} \, \mathrm{Mpc}^{-1}$ combining 42 SNe Ia with a sample of Cepheids hosted by the same galaxies. This value of $M$ is assumed to compute $\mu$ for the \textit{Pantheon +}. 
The theoretical $\mu$, assuming a flat $\Lambda$CDM model and omitting the current relativistic density parameter, negligible in the late Universe, 
is \citep{2019ApJ...875..145K}:

\begin{equation}
\begin{split}
\label{dmlcdm_corr}
& \mu_{th}   =   5 \, \mathrm{log_{10}} \, d_l \mathrm{(Mpc)} + 25 \\ &\quad  =  5 \, (1+z_{hel}) \frac{c}{H_{0}} \, \int^{z_{HD}}_{0} \frac{d z'}{\sqrt{\Omega_{M} (1+z)^{3} + (1- \Omega_{M})}} + 25
\end{split}
\end{equation}
where $d_l$ is the luminosity distance, $c$ the light speed, $\Omega_M$ the current matter density, $z_{hel}$ the redshift in the heliocentric frame, and $z_{HD}$ the redshift corrected for CMB dipole motion and peculiar velocity.
We are aware that the formula in Equation \ref{dmlcdm_corr} properly works at low redshifts, as in \citet{2019ApJ...875..145K}, where the SNe Ia sample investigated covers only $z < 0.08$, but could not be suitable for \textit{Pantheon} and \textit{Pantheon +} samples, which reach $z=2.26$. Hence, we perform all our analysis by applying both Equation \ref{dmlcdm_corr} and the same equation in which $z_{hel}$ is replaced by $z_{HD}$. The latter reads as:
\begin{equation}
\begin{split}
\label{dmlcdm}
\mu_{th}   = 5 \, (1+z_{HD}) \frac{c}{H_{0}} \, \int^{z_{HD}}_{0} \frac{d z'}{\sqrt{\Omega_{M} (1+z)^{3} + (1- \Omega_{M})}} + 25.
\end{split}
\end{equation} 
This allows us to compare the results obtained with the two different formula to compute $\mu_{th}$.
In \citet{scolnic2018} for the evaluation  of cosmological parameters the $\chi^2$ is defined as 
\begin{equation}
\label{chi2}
\chi^2=\Delta\mu^T C^{-1} \Delta\mu
\end{equation}\label{chisquare}
where $\Delta \mu = \mu_{th} - \mu_{obs}$ and $C$ is the uncertainty matrix built as $C= D_{\mathrm{stat}} + C_{\mathrm{sys}}$, with $D_{\mathrm{stat}}$ the diagonal matrix with statistical errors and $C_{\mathrm{sys}}$ the systematic covariance matrix, both provided by SNe Ia releases 
\citep{scolnic2018,2022ApJ...938..110B,pantheon+}. 
As a consequence, we are considering for each measurement its individual distance uncertainties, since different SNe Ia light-curves present different quality, signal-to-noise ratio, and contributions to the overall uncertainty, as detailed in \citet{pantheon+}. 
We here stress that we have considered both the systematic and statistical uncertainty to guarantee more reliable confidence levels of the cosmological parameters.
Indeed, the $C_{\mathrm{sys}}$ matrix is defined through a sum over the sources as follows
\begin{equation}
\label{Csys}
   C_{ij,sys} = \sum_{k=1}^{K}\Bigg(\frac{\delta\mu_i}{\delta S_k}\Bigg)\Bigg(\frac{\delta\mu_j}{\delta S_k}\Bigg) (\sigma_{S_k})^2
\end{equation}
where $k$ runs over the systematic $S_k$, $\sigma_{S_k}$ is the magnitude of each systematic error, and $\delta \mu$ is computed as the binned difference in distances when changing one of the systematic parameters.

\subsection{The Gaussianity assumption in the likelihood}
\label{normalitytests}

It is common practise to apply a Gaussian cosmological likelihood for SNe Ia. However, often the Gaussianity statement is not carefully checked each time or once the sample is enlarged.
We investigate if the argument $ \Delta \mu _{norm}  = C^{-1/2} \, \Delta \mu$ \citep{lovick2023nongaussian} of the likelihood of SNe Ia distance moduli ($\cal L$) in the \textit{Pantheon} and \textit{Pantheon +} fulfills the Gaussianity condition within a flat $\Lambda$CDM model with $\Omega_M = 0.3$ and $H_0 = 70 \, \mathrm{km} \, \mathrm{s}^{-1} \, \mathrm{Mpc}^{-1}$.
We define $\Delta \mu _{norm}$ this way to investigate the normalised residuals of the distance moduli while accounting for both the diagonal and the non-diagonal terms of the $C$ matrix, which include both statistical and systematic uncertainties. Nevertheless, we have also tested that, if we do not consider the non-diagonal terms of the $C$ matrix, thus considering only $\Delta \mu / \sigma_{\mu}$, where $\sigma_{\mu}$ is the diagonal part of $C$, our results on the Gaussianity assumption, and hence our cosmological results, do not change.
Since the computation of $\Delta \mu _{norm}$ requires the assumption of a specific cosmological model, we have chosen to report here the results obtained assuming a flat $\Lambda$CDM model with $\Omega_M = 0.3$ and $H_0 = 70 \, \mathrm{km} \, \mathrm{s}^{-1} \, \mathrm{Mpc}^{-1}$. Indeed, this value of $\Omega_M$ is the one reported in \citet{scolnic2018} and the value of $H_0 = 70 \, \mathrm{km} \, \mathrm{s}^{-1} \, \mathrm{Mpc}^{-1}$ is the one obtained in several works that combine SNe Ia with other probes \citep[e.g.][]{2020JCAP...07..045D,2021MNRAS.504..300C,2022JHEAp..34...49A}. We here notice that we have checked the impact of this cosmological assumption on our results by computing the distribution of $\Delta \mu _{norm}$ by imposing other values of $\Omega_M$. More specifically, we have tested the two extreme cases of $\Omega_M = 0.1$ and $\Omega_M = 1$. Under these new assumptions, as expected, the values of skewness and kurtosis of the distribution of $\Delta \mu _{norm}$ change. Nevertheless, both \textit{Pantheon} and \textit{Pantheon +} data still fail all the Gaussianity tests employed in our analysis and remarkably the best-fit distributions are still a logistic and a Student’s T, respectively. Hence, our results on the Gaussianity tests (Sect. \ref{normalresults}) and new best-fit likelihoods (Sect. \ref{bestfitlikelihood}), as well as the cosmological results reported in Sect. \ref{cosmologicalfits}, do not depend on the assumption of a flat $\Lambda$CDM model with $\Omega_M = 0.3$ and $H_0 = 70 \, \mathrm{km} \, \mathrm{s}^{-1} \, \mathrm{Mpc}^{-1}$.

We apply several methods to test if  $\Delta \mu _{norm}$ is normally distributed. We have applied all the tests to all individual data points of $\Delta \mu _{norm}$ and not in bins. 
$\mu_{obs}$ in Equation \eqref{muobs} is taken directly
from \citet{scolnic2018}
\footnote{\url{https://github.com/dscolnic/Pantheon}}
and \citet{pantheon+}
\footnote{\url{https://github.com/PantheonPlusSH0ES}}
, while $\mu_{th}$ is specified in Eq. \eqref{dmlcdm_corr}. When the likelihood is Gaussian there is the implicit assumption that the residual, the argument in the formula of the likelihood ($\Delta \mu _{norm}$ in Eq. \eqref{gaussianlf}), must be Gaussian. 
As anticipated, due to the weak lensing effect the distribution is not expected to be Gaussian \citep{2005ApJ...631..678H}.
Moreover, the discussion about Gaussianity in the large structure analysis has been already treated in \citet{2019MNRAS.485.2956H} along with the impact of the Gaussian assumption on cosmological parameters.
In this regard, some studies concerning to what extent the non-Gaussianity could affect the determination of the cosmological parameters have already been performed \citep{2008ApJ...678....1S,2023MNRAS.520L..68S}.
Moreover, possibilities to use non-Gaussian approaches for SNe Ia and Cepheids gathered by \citet{2016ApJ...826...56R} and Pantheon+ have been also studied in \citet{2018MNRAS.476.3861F} and \citet{2022arXiv221207917K}, respectively, but we here for the first time perform a comparative analysis with the Pantheon and Pantheon+ sample. 

We apply the Anderson-Darling \citep{stephens1976asymptotic,stephens1979tests} and Shapiro-Wilk \citep{shapiro1965analysis} normality tests, and we compute skewness and kurtosis of the distributions, the skewness \citep{doi:10.1080/00031305.1990.10475751}, the kurtosis \citep{anscombe1983distribution} and a test that combines both features \citep{d1973tests} \footnote{These analyses are performed with the scipy Python package.}.  
The Anderson-Darling and Shapiro-Wilk tests determine whether a data sample is drawn from the Gaussian distribution, and they are frequently applied in the literature (e.g. \citealt{2022ApJS..261...25D} in astrophysics and \citealt{stephens1974edf, razali2011power} in statistics).
An important property of these tests is their consistency: they can identify any small deviation
from normality, if the sample size is large enough \citep[][]{stephens1976asymptotic, Leslie_86}. Therefore, the larger the sample size, the more these tests tend to reject the null hypothesis of normal distribution \citep{yazici2007comparison}, even in case of very small deviations from Gaussianity
such as in presence of ties caused by limited measurement precision (i.e. the number of decimals). 
This weakness limits the applicability of the Anderson-Darling and Shapiro-Wilk tests 
on large samples
and must be accounted for when considering SNe Ia samples that count more than 1000 sources. 
Thus, we include investigations on skewness and kurtosis. The skewness is the third standardized moment and measures the asymmetry of the distribution as it identifies extreme values in one tail versus the other. The kurtosis is the fourth standardized moment and measures extreme values in both tails: large kurtosis corresponds to more populated tails than the ones of a Gaussian distribution, small kurtosis to less populated tails than the ones of a Gaussian distribution. 
A normal distribution has skewness and kurtosis equal to zero\footnote{We consider Fisher’s definition of kurtosis, in which 3 is subtracted from the result to give 0 for a normal distribution.}.
So, the skewness and kurtosis of a distribution provide information on the deviation from normality.
Based on this, we use the skewness and kurtosis tests, to determine if the values of skewness and kurtosis of the investigated distribution are so statistically close to zero, that it can be well approximated by a Gaussian distribution.
Furthermore, we apply the ``skewness+kurtosis" test, which contemporaneously checks departures of skewness and kurtosis from Gaussians. These complementary methods are crucial to obtain reliable results on the investigation of the Gaussianity assumption
on the SNe Ia $\Delta \mu _{norm}$.

\subsection{Fit with the new likelihoods}
\label{fit}

Since the tests described above fail (Sec. \ref{normalresults}), we further extend our analysis. First, we fit the $\Delta \mu _{norm}$ values  
to find the best-fit distribution with the Mathematica tool FindDistribution which finds the best distributions among the first 20 best-fitting distributions. We have checked the reliability of these results with python as well. More specifically, we have employed the Mathematica tool by requiring that each distribution investigated is compared with the others through all possible properties evaluated by FindDistribution, among which there are (see also Section \ref{bestfitlikelihood}) the logarithm of the likelihood value, the Bayesian Information Criterion (BIC), the Akaike Information Criterion (AIC), and the p-values of the Pearson $\chi^2$ and Cramer Von Mises tests. After this comparison, the fitting distributions are determined starting from the best-fit one to the ones that reproduce the data with highest accuracy. We report in Table \ref{tab:distributions}, the first five fitting distributions for both \textit{Pantheon} and \textit{Pantheon +} samples along with the corresponding values of the above-mentioned statistical tests. Indeed, the value of the likelihood and the BIC, and AIC criteria provide an estimate of the quality of the statistical model investigated for a given set of data. More precisely, larger values of likelihood and lower values of BIC and AIC identify the favored model. We here notice that we show in Table \ref{tab:distributions} the values of BIC and AIC computed with python. Indeed, python follows the standard definition and notation for these two criteria. This allows a more immediate and easier interpretation of the table since the preferred models are the ones with lower BIC and AIC. Additionally, the Pearson $\chi^2$ and the Cramer Von Mises tests include, respectively, the Pearson and Cramer Von Mises goodness-of-fit tests with null hypothesis that the data is drawn from a population with a specific distribution. 
For these tests, if the p-value is small enough (usually $p < 0.05$ by convention), then the null hypothesis is rejected, and we conclude that the observed data does not follow that distribution.
Thus, in these tests, the distributions with larger p-values are the preferred ones.
 The distribution here used is the first best-fit distribution shown in Table \ref{tab:distributions}. 
 
The found distribution is used in $\cal L$ ($\cal L$$_{new}$) instead of the Gaussian probability distribution function (PDF).
Thus, we fit the flat $\Lambda$CDM model with the Kelly method \citep{Kelly2007} using $\cal L$$_{new}$ and considering these cases: $H_0$ fixed and $\Omega_M$ free, $\Omega_M$ fixed and $H_0$ free, and both parameters free. 
We impose wide uniform priors $0 \leq \Omega_M \leq 1$ and $50 \, \mathrm{km} \, \mathrm{s}^{-1} \, \mathrm{Mpc}^{-1} \leq H_0 \leq 100 \, \mathrm{km} \, \mathrm{s}^{-1} \, \mathrm{Mpc}^{-1}$ to guarantee that the parameter's space is explored in all physical regions.
With the uninformative priors there is no difference on this approach and the Expectation Maximization algorithm used for the Maximum Likelihood Estimation \citep{dempster1977maximum}.
We then repeat the fit with $\cal L$$_{Gauss}$, thus comparing the outcomes.
Further, we test how the cosmological constraints vary if we change the assumptions on the fixed parameter in the cases of one free parameter. 
We consider $\Omega_M = 0.30$ and $\Omega_M = 0.34$, respectively from \citet{scolnic2018} and \citet{2022ApJ...938..110B}
for a flat $\Lambda$CDM model when considering only SNe Ia with both statistical and systematic uncertainties, as in this work.
We investigate $H_0 = 73.04 \, \mathrm{km} \, \mathrm{s}^{-1} \, \mathrm{Mpc}^{-1}$ from \citet{2022ApJ...934L...7R}, and $H_0 = 70 \, \mathrm{km} \, \mathrm{s}^{-1} \, \mathrm{Mpc}^{-1}$ based on \citet{scolnic2018} 
and chosen arbitrarily \citep{2021MNRAS.504..300C,2022JHEAp..34...49A}.

We also compare our $H_0$ with the ones from \textit{Pantheon} ($P$), \textit{Pantheon +} ($P+$), and the CMB, by computing the z-score ($\zeta$), a method to evaluate the statistical distance between a value and its reference value. $\zeta$ is defined as $\zeta= |H_{0,i} - H_{0,our}|/ \sqrt{\sigma^2_{H_{0,i}} + \sigma^2_{H_{0,our}}}$ where $i$ is one of the three $H_0$ reference values with its 1 $\sigma$ uncertainty. $H_{0,our}$ and $\sigma_{H_{0,our}}$ are the $H_0$ and its 1 $\sigma$ error obtained in our computations. In Table \ref{tab:bestfit} $\zeta_{CMB}$ is calculated referred to $H_0 = 67.4 \pm 0.5 \, \mathrm{km} \, \mathrm{s}^{-1} \, \mathrm{Mpc}^{-1}$, $\zeta_P$ compared to $H_0 = 70.00 \pm 0.13 \, \mathrm{km} \, \mathrm{s}^{-1} \, \mathrm{Mpc}^{-1}$, and $\zeta_{P+}$ with respect to $H_0 = 73.04 \pm 1.04 \, \mathrm{km} \, \mathrm{s}^{-1} \, \mathrm{Mpc}^{-1}$.
The parameter $\zeta$ helps to quantify the discrepancy between the $H_0$ values obtained in our analyses and the ones assumed as references.
Additionally, we use the $\zeta$ parameter also to compare the values of $\Omega_M$ and $H_0$ obtained with $\mathcal{L} _{new}$ and $\mathcal{L}_{Gauss}$ for both \textit{Pantheon} and \textit{Pantheon +} samples. Following the previous definition of $\zeta$, we define $\zeta_{\mathcal{L},\Omega_M}$ and $\zeta_{\mathcal{L},H_0}$ as $\zeta_{\mathcal{L},\Omega_M}= |\Omega_{M,\mathcal{L}_{new}} - \Omega_{M,\mathcal{L}_{Gauss}}|/ \sqrt{\sigma^2_{\Omega_{M,\mathcal{L}_{new}}} + \sigma^2_{\Omega_{M,\mathcal{L}_{Gauss}}}}$ and $\zeta_{\mathcal{L},H_0}= |H_{0,\mathcal{L}_{new}} - H_{0,\mathcal{L}_{Gauss}}|/ \sqrt{\sigma^2_{H_{0,\mathcal{L}_{new}}} + \sigma^2_{H_{0,\mathcal{L}_{Gauss}}}}$, where the subscripts $\mathcal{L}_{new}$ and $\mathcal{L}_{Gauss}$ refer to the specific likelihood used to fit the cosmological parameters. The obtained $\zeta_{\mathcal{L},\Omega_M}$ and $\zeta_{\mathcal{L},H_0}$ are reported in Table \ref{tab:zscore:likelihood} for each case analysed in this work. 

Since the z-score allows us to quantitatively compare two results and evaluate their relative compatibility or discrepancy, we use this parameter also to test if the cosmological results obtained in all cases by employing Equation \ref{dmlcdm_corr} are consistent with the ones derived by applying Equation \ref{dmlcdm}. Thus, we define $\zeta_{\Omega_M}$ and $\zeta_{H_0}$ as $\zeta_{\Omega_M}= |\Omega_{M,2} - \Omega_{M,3}|/ \sqrt{\sigma^2_{\Omega_{M,2}} + \sigma^2_{\Omega_{M,3}}}$ and $\zeta_{H_0}= |H_{0,2} - H_{0,3}|/ \sqrt{\sigma^2_{H_{0,2}} + \sigma^2_{H_{0,3}}}$, where the subscripts 2 and 3 refer to the Equation \ref{dmlcdm_corr} and $\ref{dmlcdm}$ used respectively to fit the cosmological parameters. The obtained $\zeta_{\Omega_M}$ and $\zeta_{H_0}$ are reported in Table \ref{tab:zscore} for each case studied in this work. More specifically, in this table we present the sample (i.e. \textit{Pantheon} or \textit{Pantheon +}) in the first column, the cosmological case considered (i.e. if $H_0$ is fixed to 70 or 73.04, if $\Omega_M$ is fixed to 0.3 or 0.34, or if both $\Omega_M$ and $H_0$ are free to vary) in the second column, and the values of $\zeta_{\Omega_M}$ and $\zeta_{H_0}$ in the other columns. The table is also divided to distinguish the two different likelihoods adopted. The symbol "-" refers to the cases in which $H_0$ is fixed and thus $\zeta_{H_0}$ cannot be computed and the cases in which $\Omega_M$ is fixed and, as a consequence, $\zeta_{\Omega_M}$ cannot be calculated.

\section{Results}
\label{results}

\subsection{Gaussianity tests results}
\label{normalresults}

Both SNe Ia samples do not pass neither the Anderson-Darling nor the Shapiro-Wilk tests.
Thus, the null hypothesis that $\Delta \mu _{norm}$ comes from a Gaussian distribution should be rejected. 
The Anderson-Darling test rejects the Gaussianity hypothesis to a significance level $> 15\%$. This rejection is confirmed by the Shapiro-Wilk test as its p-value is $< 5\%$, the minimum value required to accept the null hypothesis.
Given the limits of these methods, we compute the skewness and kurtosis of the $\Delta \mu$ normalized distributions. The \textit{Pantheon} sample has skewness = -0.08 and kurtosis = 0.44 corresponding to an excess of the tail on the left, while the \textit{Pantheon +} presents the skewness =-0.50 and more prominent tails with 
kurtosis = 3.9 (see Fig. \ref{fig:hist}). 
As also visible from the right panel of Fig. \ref{fig:hist}, the skewness of \textit{Pantheon +} is larger than the one of \textit{Pantheon}.
Neither of the samples passes the skewness, kurtosis and ``skewness+kurtosis" tests.
As anticipated in Sect. \ref{normalitytests}, we also verify that the results of all the normality tests are independent on the assumed cosmology and they do not change when using Equation \ref{dmlcdm} in place of Equation \ref{dmlcdm_corr}.
All these independent normality tests result in coherent and consistent results guaranteeing the common outcome and strongly proving that the $\Delta \mu _{norm}$ of SNe Ia is not normally distributed, neither for the \textit{Pantheon} nor for the \textit{Pantheon +} sample.
The Gaussianity implicit assumption for SNe Ia $\Delta \mu _{norm}$ 
is not statistically legitimated and it should be checked for each SNe Ia sample used. 

\subsection{New best-fit distributions}
\label{bestfitlikelihood}

\begin{figure*}
\centering
    \includegraphics[width=.33\textwidth]{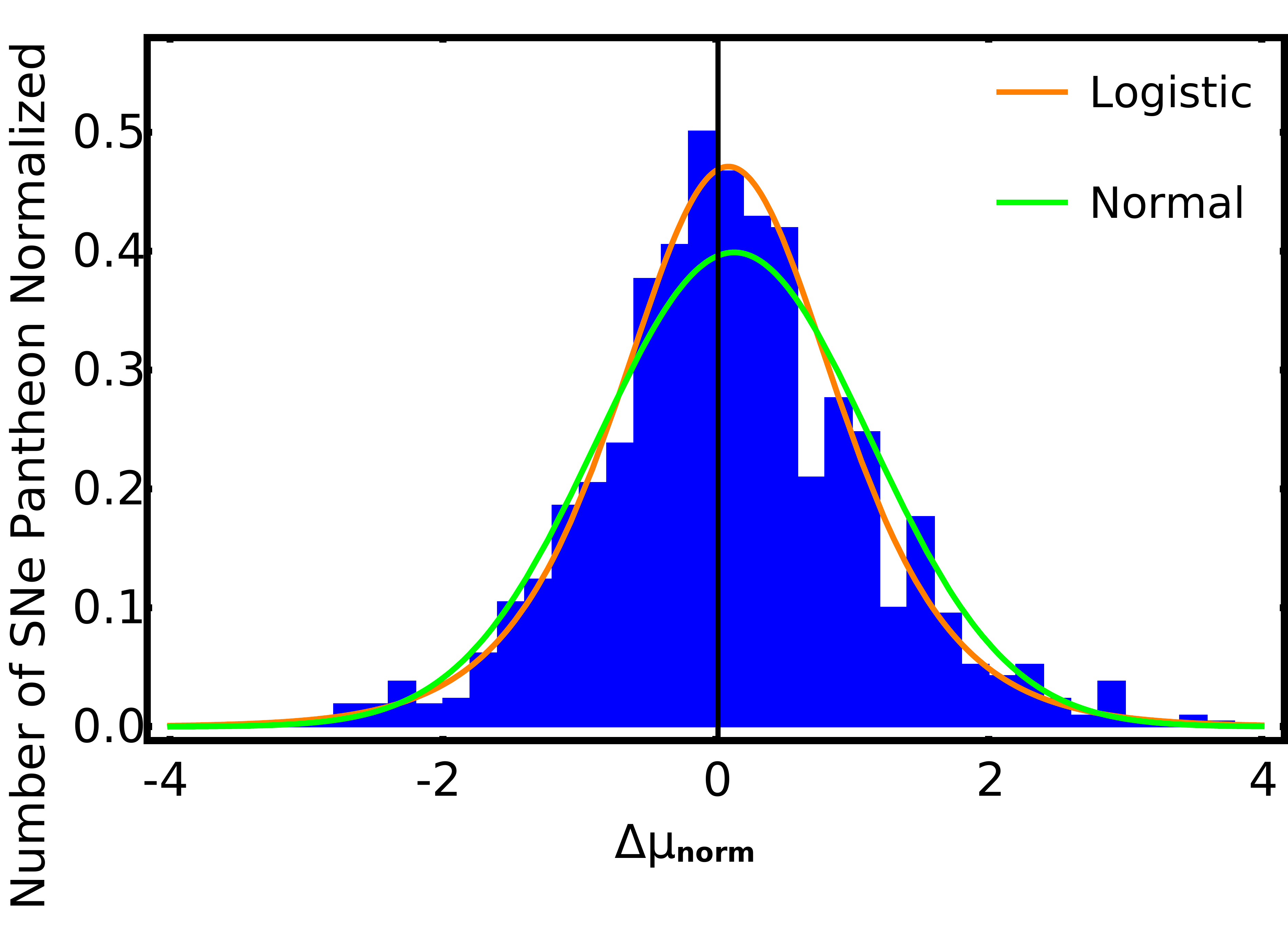}    \includegraphics[width=.33\textwidth]{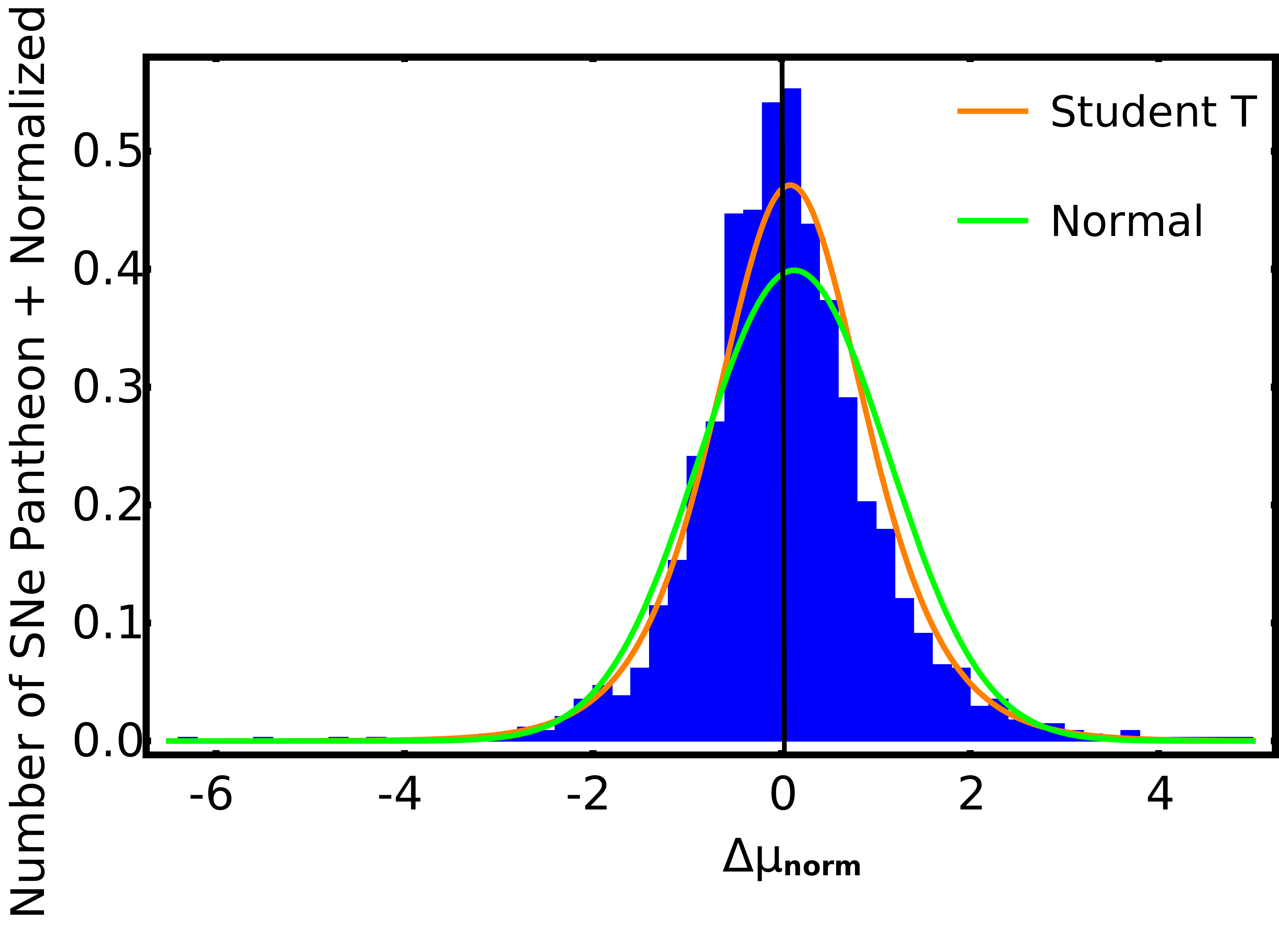}
    \includegraphics[width=.33\textwidth]{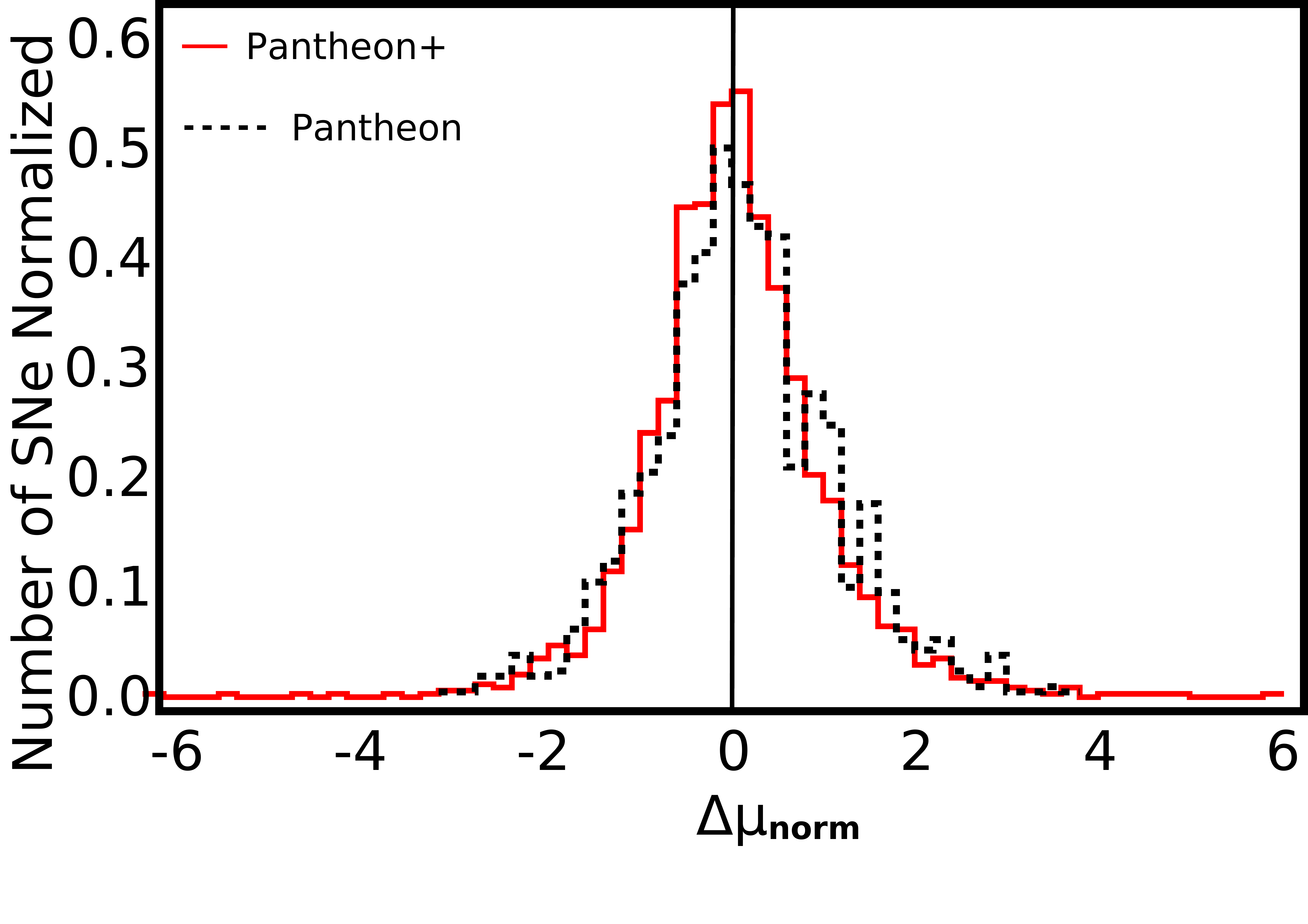}
    \caption{ Normalized $\Delta \mu _{norm}$ histogram, defined as $\Delta \mu _{norm}  = C^{-1/2} \, \Delta \mu$, for the 1048 SNe Ia in \textit{Pantheon} (left panel) and the 1701 SNe Ia in \textit{Pantheon +} (middle panel). The green curve is the best-fit Gaussian distribution, while the orange curves are the best-fit logistic (left panel) and Student's t (middle panel) distributions. Right panel shows the superimposition of the \textit{Pantheon} and \textit{Pantheon +} distributions. In all panels the vertical black line marks the zero line.}
    \label{fig:hist}
\end{figure*}

Thus, the $\Delta \mu _{norm}$ distributions are not well-fitted by the Gaussian PDF, defined as:

\begin{equation}
\label{gaussianlf}
\displaystyle
\mathrm{PDF_{Gaussian}} = \frac{1}{\sqrt{2 \, \pi} \, \sigma} \, e^{-\frac{1}{2} \left(\frac{x- \hat{x}}{\sigma}\right)^2}=\frac{1}{\sqrt{2 \, \pi} \, \sigma} \, e^{-\frac{1}{2} \chi^2}
\end{equation}
where $\hat{x}$ is the mean, $\sigma$ is the standard deviation, $x -\hat{x}=\Delta \mu$ and $\chi^2$ is the same defined in Eq. \eqref{chi2}.
Hence, we fit the $\Delta \mu$ histograms (see Sec. \ref{normalitytests}) obtaining as the best fit for the \textit{Pantheon} sample a logistic distribution, with the following PDF
\begin{equation}
\label{logistic}
\mathrm{PDF_{logistic}} = \frac{e^{-\frac{(x-\hat{x})}{s}}}{s \, \left(1+ e^{\frac{-(x-\hat{x})}{s}}\right)^2} = \frac{e^{- \chi'^2}}{s \, \left(1+ e^{- \chi'^2}\right)^2}
\end{equation}
where  $s$ is scale and the variance $\sigma^2$ is $\sigma^2 = (s^2 \, \pi^2)/3$. As for the Gaussian PDF, we consider the whole covariance matrix with both diagonal and non-diagonal components to account for all contributions to the uncertainty. Indeed, in the case of the logistic distribution, $\chi^2$ is the same defined in Eq. \eqref{chi2} but in which the elements of $C$ are multiplied by $3 \, \sigma^2 / \pi^2$. Hence, $\chi ^{'2}$ is defined as $\chi^{'2}=\Delta\mu^T (3 \, \sigma ^2 \, C \, / \pi^2)^{-1} \Delta\mu$.
Left panel of Fig. \ref{fig:hist} shows the best-fit distributions normalized with the uncertainties on the singular SNe Ia of the logistic with $\hat{x}= -0.019$ and $s=0.55$ (orange) and the Gaussian with $\hat{x}=0.030$ and $\sigma= 0.99$ (green). 

The same analysis is then repeated for the \textit{Pantheon +} data. In this case, the distribution that best fits the histogram of $\Delta \mu _{norm}$ is a Student's t-distribution, for which the generalized PDF is given by
\begin{equation}
\begin{split}
\label{student}
 \mathrm{PDF_{student}} & = \frac{\Gamma\left(\frac{\nu +1}{2}\right)}{\sqrt{\nu \, \pi} \, s \, \Gamma \left(\frac{\nu}{2}\right)} \, \left[1 + \frac{((x- \hat{x})/s)^2}{\nu}\right]^{-\frac{\nu +1}{2}}  \\ & \quad = \frac{\Gamma\left(\frac{\nu +1}{2}\right)}{\sqrt{\nu \, \pi} \, s \, \Gamma \left(\frac{\nu}{2}\right)} \, \left[1 + \frac{\chi^{'2}}{\nu}\right]^{-\frac{\nu +1}{2}}
\end{split}
\end{equation}
where $\Gamma$ is the gamma function, $\nu$ are the degrees of freedom, and $\sigma^2 = (s^2 \, \nu) / (\nu -2)$. As explained for the logistic case, this relation between $s$ and $\sigma$ allows us to substitute $C$ used in $\chi^2$ of Eq. \eqref{chi2} for the Student's t. More specifically, $\chi'^2=\Delta\mu^T \left[(\nu -1) \, \sigma ^2 \, C \, / \nu \right]^{-1} \Delta\mu$.
In our case, $x=\Delta \mu$ and $\hat{x}=0$, as before.
The best-fit distributions of the Student's t with $\hat{x}= 0.01$, $s=0.20$, and $\nu = 4.2$ (orange) and the Gaussian with $\hat{x}=0.018$ and $\sigma= 0.30$ (green) are shown in the middle panel of Fig. \ref{fig:hist} where the distribution is normalized with the SNe Ia uncertainties (Eq. \eqref{gaussianlf}).
For the \textit{Pantheon} and \textit{Pantheon +} respectively, the logistic and the Student's t distributions better reproduce $\Delta \mu _{norm}$ in all the features (e.g. peak, central width, and tails) compared to the Gaussian PDF.
We have also proved that these results remain the same if we apply Equation \ref{dmlcdm} instead of Equation \ref{dmlcdm_corr} and if we change the cosmological assumption of a flat $\Lambda$CDM model with $\Omega_M = 0.3$ and $H_0 = 70 \, \mathrm{km \, s^{-1} \, Mpc^{-1}}$, as already stated in Sect. \ref{normalitytests}.
The first best-fit distribution for both \textit{Pantheon} and \textit{Pantheon +} samples is reported along with the other four that better fit the data in Table \ref{tab:distributions}. As anticipated in Sect. \ref{fit}, the most appropriate, namely the best-fit distributions, are ranked through the evaluation of the logarithm of the likelihood, BIC and AIC parameters, and the p-values of Pearson $\chi^2$ and Cramer Von Mises tests.
We have also proved that the logistic for the \textit{Pantheon} and the Student-T distribution for the \textit{Pantheon +} are favoured against the Gaussian distributions with the BIC. We used for comparison the table in \citet{kass1995bayes} to distinguish among the most favoured models. We have used the definition of $2\mathrm{log}_e(B_{10})$, where the Bayes factor $B_{10}=e^{((BIC_{0}-BIC_1)/2)}$ \citep{wagenmakers2007practical}, $B_0$ is the reference model (the Gaussian), and $B_1$ is the alternative model (either the logistic or the T-student distribution). Thus, $2\mathrm{log}_e(B10)=BIC_{0}-BIC_1=\Delta_{BIC}$. We found $\Delta_{BIC,Pantheon}=7$ showing strong evidence for the Logistic against the Gaussian distribution, and $\Delta_{BIC,Pantheon+}=150$ showing very strong evidence for the T-Student against the Gaussian for the Pantheon+. By definition, for $6 \leq \Delta_{BIC} \leq 10$ the evidence against the null hypothesis (the Gaussian being the favorite the model) is strong, when $\Delta_{BIC}>10$ the evidence against the null hypothesis is very strong.
However, we would like to stress that the BIC works under the assumption that the variables are independent. Since SNe Ia are correlated, the BIC test may less trustworthy. Thus, we prefer to apply the AIC, since it does not include the number of SNe Ia. Indeed, it is also worth noting that some of the SNe Ia are repeated in the sample, since they belong to different survey. When computing the AIC, we still find that the logistic distribution is favored against the Gaussian since the AIC of the logistic is lower than the one of the Gaussian of a factor 8. Moreover, also the Student-T is favored against the Gaussian, with an AIC parameter that is lower of a factor 150 compared to the one of the Gaussian distribution. 
However, we would like to take a further step in the discussion. The number of actual SNe Ia observed is smaller than the total number of SNe Ia either in the Pantheon or in the Pantheon+ sample, thus the BIC criterion could be corrected for a smaller number of SNe Ia. When the number of parameters is different, e.g. the T-student has 3 parameters instead of two of the Gaussian, the difference of the BIC would be even larger, thus showing the reliability of these likelihoods against the Gaussian one. Although there may be more refined criteria for this study, the analysis of this comparison would require a more complex statistical study, which goes far beyond the scope of the paper.
 Moreover, we have proved the goodness of the fits with the Anderson-Darling one-sample test. The results yield a probability value that the fit is not randomly drawn, $p=0.97$ for the \textit{Pantheon} sample for the logistic distribution, and $p=0.88$ for the \textit{Pantheon +} for the Student-T distribution.
 For visual clarity, right panel of Fig. \ref{fig:hist} shows the comparison between the normalized residuals $\Delta \mu _{norm}$ of Pantheon (in dashed gray) and Pantheon+ (in continuous red). This superimposition clearly proves that the Pantheon+ distribution is more peaked and presents heavier tails compared to Pantheon. This difference stems from the sources present in Pantheon+ and not in Pantheon. Furthermore, this discrepancy causes the difference between the two corresponding best-fit distributions, namely the change from the logistic distribution for the Pantheon SNe Ia to the Student-T for the Pantheon+.
 Although our analysis is statistically reliable, we here stress that this reasoning is valid if the distributions of Fig. \ref{fig:hist} are derived from a perfect sample and modeling of selection effects. If, however, there are problems with some of these sources, such as bad background subtraction, cosmic ray hit etc. this surely can impact the distribution, but a deep analysis of these sources is far beyond the scope of the current analysis. We here assume that the Pantheon and Pantheon+ samples are trustworthy sources.

\begin{table*}
\centering
\begin{tabular}{ccccccc}
\hline
\textit{Pantheon} & Distributions & Log Likelihood & AIC & BIC & Pearson $\chi^2$ & Cramer Von Mises\\
\hline
 & Logistic & 0.54 & 51.5 & -7555 & 0.84 & 0.71   \\
 & Gaussian  & 0.53 & 53.3 & -7030 & 0.006 & 0.006   \\
 &  Student's t  & 0.53 & 47.6 & -6820 & 0.02 & 0.0001  \\
 &  Weibull  & 0.52 & 59.7 & -6577 & 0.001 & 0.0001 \\
 & Laplace & 0.51  & 73.2 & -6764 & 2 x $10^{-5}$ & 1 x $10^{-6}$ \\
\hline
\textit{Pantheon +} & \\
\hline
 &  Student's t  & 0.33 & 176.1 & -16890 & 0.08 & 0.06\\
 &  Logistic  & 0.33 & 218.6 & -15201 & 0.08 & 0.6 \\
 &  Laplace  & 0.30 & 178.9 & -12943 & 3 x $10^{-10}$ & 0.003 \\
 &  Gaussian  & 0.28 & 334.3 & -12567 & 8  x $ 10^{-23}$ & $9 $ x $ 10^{-10}$\\
 &  Cauchy  & 0.20 & 125.5 & -12005 & 7 x $10^{-53}$ & 2 x $10^{-5}$\\
\hline
\end{tabular}
\caption{First five best-fit distributions of $\Delta \mu _{norm}$ for both \textit{Pantheon} and \textit{Pantheon +} samples. The columns next to each distribution show the values of the following computed parameters: logarithm of the likelihood, AIC, BIC, p-value of Pearson $\chi^2$ test, and p-value of the Cramer Von Mises test (see also Sect. \ref{fit}).}
\label{tab:distributions}
\end{table*}

\subsection{Cosmology with the Gaussian and new likelihoods}
\label{cosmologicalfits}
 
We here comment and interpret our cosmological results. As shown in Table \ref{tab:zscore}, the values obtained for $\Omega_M$ and $H_0$ when applying Equation \ref{dmlcdm_corr} and Equation \ref{dmlcdm} are completely consistent within $0.3 \sigma$, thus proving that our results are robust and independent on the specific formula used to compute the theoretical distance moduli. Hence, the following considerations are valid for both cases.

\begin{figure*}
\centering
    \includegraphics[width=.49\textwidth]{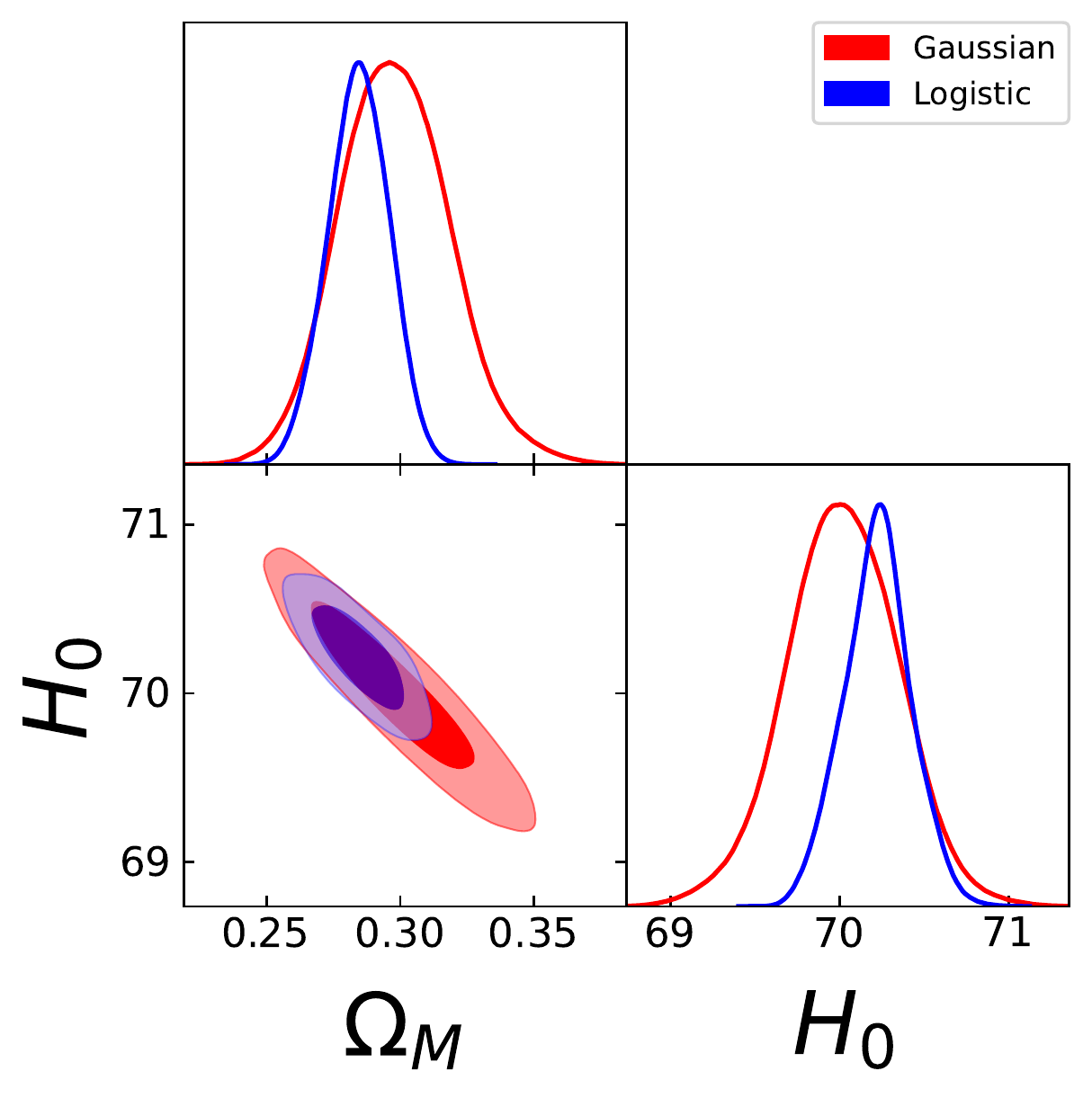}    \includegraphics[width=.49\textwidth]{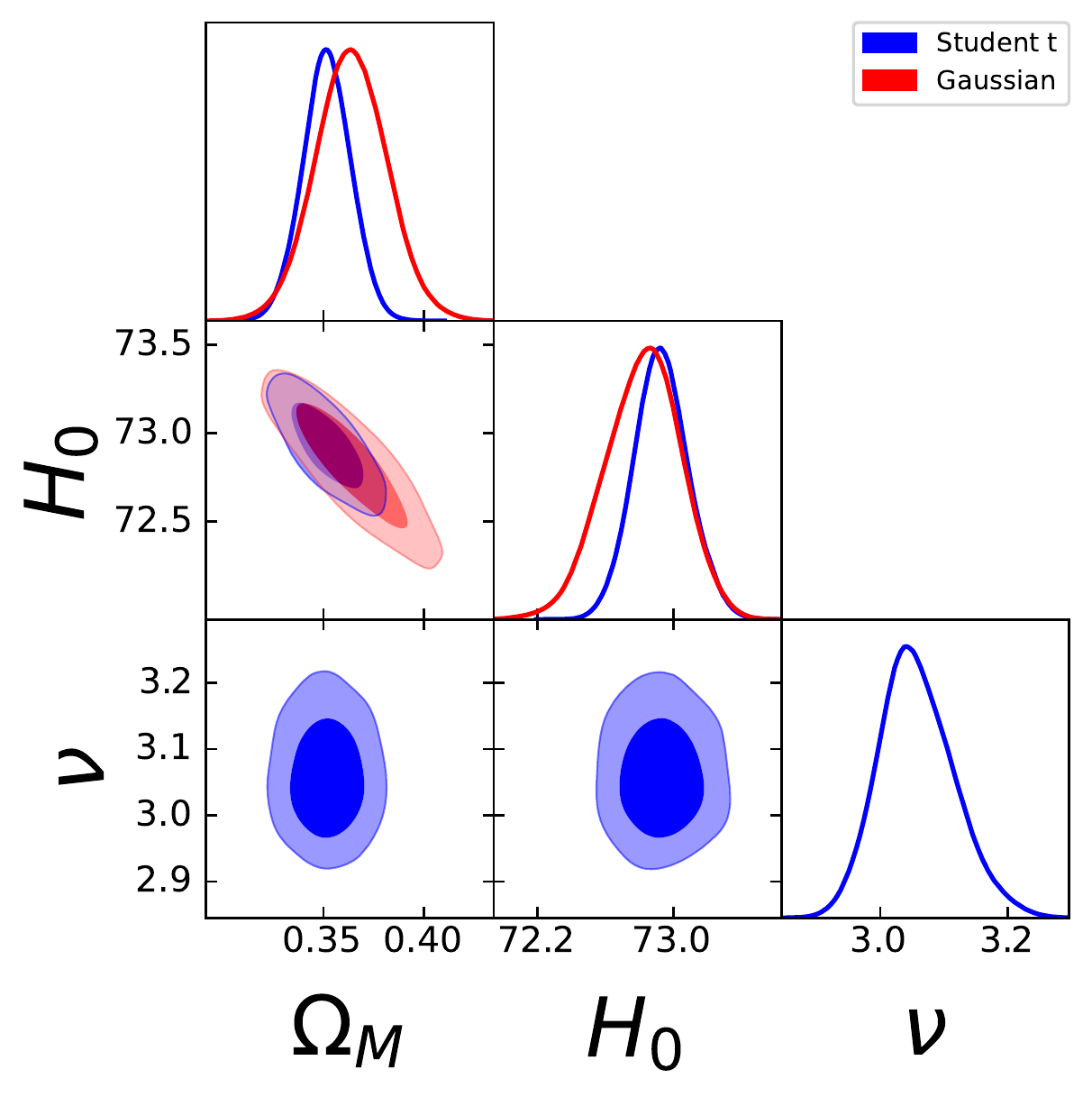}
    
    \caption{Fit of the flat $\Lambda$CDM model with $\Omega_M$ and $H_0$ free parameters. Left panel shows the results for \textit{Pantheon} SNe Ia with both $\cal L$$_{Gauss}$ and $\cal L$$_{logistic}$ as in the legend. Right panel shows the contours for the \textit{Pantheon +} sample with both $\cal L$$_{Gauss}$ and $\cal L$$_{Student}$ as illustrated in the legend.}
\label{fig: Om+H0}
\end{figure*}

Using $\cal L$$_{new}$, we fit the flat $\Lambda$CDM model on the \textit{Pantheon} and \textit{Pantheon+} samples (see Section \ref{fit}). Table \ref{tab:bestfit} reports the best-fit values with 68\% confidence level. Left panel of Fig. \ref{fig: Om+H0} show the corner plots (with blue contours) from the logistic likelihood ($\cal L$$_{logistic}$), while right panel of Fig. \ref{fig: Om+H0} the ones from the Student's t likelihood ($\cal L$$_{Student}$) in blue. For the Student's t, we treat the degrees of freedom $\nu$ as a free parameter by imposing a wide uniform prior in the range $0 < \nu <6$. In principle, we could consider it as a free parameter or we could fix it a priori and fit only the cosmological parameters. To investigate this choice, we have tested both approaches. This analysis has shown that, if we fix $\nu$ to its best-fit value $\nu = 4.2$ obtained from the fit of the histogram of $\Delta \mu _{norm}$ for \textit{Pantheon+} in the middle panel of Fig. \ref{fig:hist} (see Section \ref{normalresults}), the values of $\Omega_M$ and $H_0$ obtained are compatible within $0.2 \sigma$ with the ones derived when $\nu$ is free to vary. Thus, this choice does not impact our cosmological results. Moreover, $\nu$ does not manifest any correlation with the cosmological free parameters, as can be seen in right panel of Fig. \ref{fig: Om+H0}, hence the estimates of $\Omega_M$ and $H_0$ are independent on the treatment of $\nu$, which is if we fix it or leave it free. Thus, we have decided to apply the more general approach of using $\nu$ as a free parameter, which also allows us to perform a more accurate analysis, without the need for knowing the value of this parameter a priori. 
The results of the fit of the flat $\Lambda$CDM model with $\cal L$$_{Gauss}$ are in Table \ref{tab:bestfit} and
in red in Fig. \ref{fig: Om+H0}.

Considering the \textit{Pantheon} sample, the cosmological parameters obtained with $\cal L$$_{Gauss}$ and $\cal L$$_{logistic}$ are compatible in 1 $\sigma$, except for $\Omega_M$ when $H_0 = 73.04 \, \mathrm{km} \, \mathrm{s}^{-1} \, \mathrm{Mpc}^{-1}$, that shows a $2-3 \, \sigma$ discrepancy. 
Concerning the uncertainties,
$\cal L$$_{logistic}$ shows significantly tighter constraints when both $\Omega_M$ and $H_0$ are free parameters: the errors on $\Omega_M$ and $H_0$ are reduced by $43 \%$ (from 0.021 to 0.012) and $41 \%$ (from 0.34 to 0.20), respectively. Same conclusions are drawn for the \textit{Pantheon+}. The only difference is that
the $3 \, \sigma$ discrepancy in $\Omega_M$ emerges when $H_0 = 70  \, \mathrm{km} \, \mathrm{s}^{-1} \, \mathrm{Mpc}^{-1}$.
This can be ascribed to the $H_0 = 73.04  \, \mathrm{km} \, \mathrm{s}^{-1} \, \mathrm{Mpc}^{-1}$ assumed for \textit{Pantheon+}.
The use of $\cal L$$_{Student}$ reduces the errors on $\Omega_M$ and $H_0$, when both are free parameters, by $42 \%$ (from 0.019 to 0.011) and $33 \%$ (from 0.24 to 0.16), respectively.
Further, \citet{planck2018} reported $\Omega_M = 0.315 \pm 0.007$. Thus, the discrepancy with the values in Table \ref{tab:bestfit}, when $\Omega_M$ and $H_0$ are free, increases using $\cal L$$_{new}$ due to the reduced errors. Quantitatively, $\cal L$$_{logistic}$ shows a 2 $\sigma$ discrepancy compared to 1 $\sigma$ of $\cal L$$_{Gauss}$, while $\cal L$$_{Student}$ increases the discrepancy from 2 to 3 $\sigma$.

The current study highlights differences in the uncertainties of the Hubble constant ($H_0$) as compared to the findings reported in previous works by \citet{2019ApJ...876...85R} and \citet{2022ApJ...934L...7R}, which focused on the \textit{Pantheon} and \textit{Pantheon+} data sets.
As already detailed in \citet{Bargiacchi2023MNRAS.521.3909B}, these variations in uncertainties are attributed to distinct approaches used in determining $H_0$ in the different studies. Specifically, \citet{2019ApJ...876...85R} and \citet{2022ApJ...934L...7R} employed various methods, such as analyzing Cepheids in SNe Ia hosts, Cepheids as anchors or non-SNe Ia hosts, SNe Ia in Cepheids hosts, external constraints, and SNe Ia in the Hubble flow. In their analyses, they also considered free parameters in the fit, including $5 \, \mathrm{log}H_0$, from which $H_0$ and its uncertainty were derived. Additionally, a contribution to the systematic uncertainty from different analysis variants was included.
In \citet{2022ApJ...934L...7R}, the value of $H_0$ was used, along with its corresponding $M$ parameter, to calculate the distance moduli and their uncertainties for the \textit{Pantheon+} release. In contrast, the \textit{Pantheon} release assumed $H_0=70$ when providing distance moduli.
In this present study, a distinct approach is taken. The entire SNe Ia sample (1048 sources for \textit{Pantheon} and 1701 sources for \textit{Pantheon+}) is considered, and direct fitting of the distance moduli and their uncertainties, is performed contrary to the approach in Riess in which the fit is calibrated with 42 Cepheids.
In our approach, the values of $M$ are fixed in the fitting process, to the corresponding values of the distance moduli to either $H_0=70$ or $H_0=73.04$, depending on the data set (\textit{Pantheon} or \textit{Pantheon +}), respectively.
As a result of this approach, the uncertainties on $H_0$ obtained in the current study are significantly reduced compared to those reported in \citet{2019ApJ...876...85R} and \citet{2022ApJ...934L...7R}.

Additional interesting trends are revealed when comparing the cosmological results from the two SNe Ia samples, independently on the $\cal L$ considered. With \textit{Pantheon}, we recover $\Omega_M = 0.3$ and $H_0 = 70  \, \mathrm{km} \, \mathrm{s}^{-1} \, \mathrm{Mpc}^{-1}$, as in \citet{scolnic2018}, except when assuming $\Omega_M =0.34$ and $H_0 = 73.04  \, \mathrm{km} \, \mathrm{s}^{-1} \, \mathrm{Mpc}^{-1}$. 
Conversely, with \textit{Pantheon+}, we recover $\Omega_M = 0.35$ and $H_0 = 73 \, \mathrm{km} \, \mathrm{s}^{-1} \, \mathrm{Mpc}^{-1}$ as in \citet{2022ApJ...938..110B} and \citet{2022ApJ...934L...7R}, except when fixing $H_0 = 70 \, \mathrm{km} \, \mathrm{s}^{-1} \, \mathrm{Mpc}^{-1}$.
The significant shifts in $\Omega_M$
show how strongly the choice of 
$M$ (and thus $H_0$) used to compute $\mu$ affects the cosmological results.
We have performed reverse engineering reasoning: we have used the same distance moduli provided by \citet{scolnic2018} and \citet{2022ApJ...938..110B} by using the covariance matrix which include both statistics and systematic uncertainties.
 We have used the same assumption on $H_0$ used in \citet{scolnic2018} and in \citet{2022ApJ...938..110B}. Thus, if there was no hidden selection selection biases or redshift evolution at play, we should have recovered the values of the assumption for $\Omega_M$. However, when we fix this value of $H_0$ we have a varying $\Omega_M$ as an inherent evidence of an evolving trend in $H_0$ already extensively discussed in \citet{2021ApJ...912..150D,2022Galax..10...24D} for which \citet{Schiavone:2022wvq} have further provided a theoretical explanation.
This trend of $\Omega_M$ (also discussed in \citet{2022arXiv220611447C,2022arXiv221102129C}) can be either the consequence that we are forcing the value of $H_0$ to be fixed while an evolving trend is present in the data or that there is an inherent crisis of the $\Lambda$CDM model that brings to an evolving trend of $\Omega_M$. 
Another possibility is that there are selection biases or evolutionary effects at play which are still unclear. 
The difference in the trend of $\Omega_M$ and $H_0$ could be due to the fact that in the local Universe we do not see a dependence on $\Omega_M$ and $H_0$ which instead is revealed at high-$z$. Anyhow, regardless of the physical meaning and the explanation such a trend is revealed in the data and is important to highlight the impact associated with the quantification of the $H_0$ tension.
The evolving trend of $H_0$ is shown also in \citet{2022arXiv221200238J} with $H(z)$ and additional data samples for the SNe Ia.


Additionally, we compute $\zeta$ (see Sec. \ref{fit}) for each cosmological computation (Table \ref{tab:bestfit}).
The high $\zeta_P$ values ($\sim 20$) when calculated for the \textit{Pantheon+} sample may depend on the smaller error on the reference value $H_0 = 70.00 \pm 0.13 \, \mathrm{km} \, \mathrm{s}^{-1} \, \mathrm{Mpc}^{-1}$ compared to the errors of the reference values pertinent to $\zeta_{CMB}$ and $\zeta_{P+}$.
$\zeta_{P+}$ is always $\sim 3$ when computed for \textit{Pantheon}; $\zeta_{CMB}$ is higher for \textit{Pantheon+} ($\zeta_{CMB} \sim 12$) than \textit{Pantheon} ($\zeta_{CMB} \sim 5$) as the $H_0$ value in \citet{2022ApJ...934L...7R} is more discrepant from the one derived from CMB compared to the $H_0$ in \citet{scolnic2018}.
Concerning the $\zeta$ reported in Table \ref{tab:zscore:likelihood} and computed to compare the cosmological parameters obtained with $\mathcal{L}_{new}$ and $\mathcal{L}_{Gauss}$, we notice that the highest values are the ones obtained when $H_0$ is fixed and $\Omega_M$ is free to vary. Indeed, for \textit{Pantheon} the maximum value is $\zeta_{\mathcal{L}, \Omega_M} = 3.1$, obtained when $H_0 = 73.04 \, \mathrm{km} \, \mathrm{s}^{-1} \, \mathrm{Mpc}^{-1} $, while for \textit{Pantheon +} we obtain the maximum value of $\zeta_{\mathcal{L}, \Omega_M} = 3.7$ when $H_0 = 70 \, \mathrm{km} \, \mathrm{s}^{-1} \, \mathrm{Mpc}^{-1} $. This result is expected since \textit{Pantheon} SNe Ia favour $H_0 = 70 \, \mathrm{km} \, \mathrm{s}^{-1} \, \mathrm{Mpc}^{-1} $, while \textit{Pantheon +} prefers $H_0 = 73.04 \, \mathrm{km} \, \mathrm{s}^{-1} \, \mathrm{Mpc}^{-1} $. The reason why we do not obtain high values also of $\zeta_{\mathcal{L}, H_0}$ when $\Omega_M$ is fixed is that the uncertainties on $H_0$ are larger than the ones on $\Omega_M$ (see Table \ref{tab:bestfit}) and hence the denominator in the definition of the $\zeta_{\mathcal{L}, H_0}$ is larger than the one of $\zeta_{\mathcal{L}, \Omega_M}$ making $\zeta_{\mathcal{L}, H_0}$ smaller than $\zeta_{\mathcal{L}, \Omega_M}$. 
The fact that the values of $\zeta$ reported in Table \ref{tab:zscore:likelihood} are always $< 1$, except for the two cases discussed above, shows that the best-fit values of $\Omega_M$ and $H_0$ obtained with the new best-fit likelihoods and the Gaussian likelihood are compatible, as already stressed above. Nonetheless, the crucial point is that the use of $\mathcal{L}_{new}$ significantly reduces the uncertainties on these cosmological parameters compared to $\mathcal{L}_{Gauss}$. Indeed, $\cal L$$_{logistic}$ reduces the uncertainties on $\Omega_M$ and $H_0$ up to $ 43 \%$ (from 0.021 to 0.012) and $41 \%$ (from 0.34 to 0.20), respectively, while $\cal L$$_{Student}$ reduces the error on $\Omega_M$ and $H_0$ up to $42 \%$ (from 0.019 to 0.011) and $33 \%$ (from 0.24 to 0.16), respectively. 


\begin{table*}
\centering
\resizebox{\linewidth}{!}{%
\begin{tabular}{cccccccccccc}
\hline
\multicolumn{7}{c}{$\cal L$$_{logistic}$}&\multicolumn{5}{c}{$\cal L$$_{Gauss}$}\\
\hline
\textit{Pantheon} & $H_0$ & $\Omega_M$ & & $\zeta_{CMB}$ & $\zeta_{P}$ & $\zeta_{P+}$ & $H_0$ & $\Omega_M$ & $\zeta_{CMB}$ & $\zeta_{P}$ & $\zeta_{P+}$\\
\hline
 & \textbf{70} & $0.295 \pm 0.007$ & & & & & \textbf{70} & $0.300 \pm 0.009$ & & &\\
 & \textbf{73.04} & $0.173 \pm 0.006$ & & & & & \textbf{73.04} & $0.144 \pm 0.007$ & & & \\
 & $70.00 \pm 0.13$ & \textbf{0.3} & & 5.73 & 0 & 2.90 & $70.00 \pm 0.13 $ &  \textbf{0.3} & 5.73 & 0 & 2.90 \\
 & $69.47 \pm 0.12$ & \textbf{0.34} & &4.73 & 2.96 &3.41 & $69.43 \pm 0.14$ & \textbf{0.34} & 4.60 &2.98 & 3.44\\
 & $70.21 \pm 0.20$ &$0.285 \pm 0.012$ & &5.89 & 0.88 &2.67 & $69.99 \pm 0.34$ & $0.300 \pm 0.021$ &4.88 & 0.03 & 2.79\\
\hline
\multicolumn{7}{c}{$\cal L$$_{Student}$}&\multicolumn{5}{c}{$\cal L$$_{Gauss}$}\\
\hline
\textit{Pantheon+} & $H_0$ & $\Omega_M$ & $\nu$ & $\zeta_{CMB}$ & $\zeta_{P}$ & $\zeta_{P+}$ & $H_0$ & $\Omega_M$ & $\zeta_{CMB}$ & $\zeta_{P}$ & $\zeta_{P+}$
\\
\hline

 & \textbf{70} & $0.532 \pm 0.010$ & $3.35 \pm 0.07$ & & & & \textbf{70} & $0.587 \pm 0.011$ & & &  \\
 & \textbf{73.04} & $0.345 \pm 0.007$ & $3.05 \pm 0.06$ & & & & \textbf{73.04} & $0.349 \pm 0.010$ & & & \\
 & $73.50 \pm 0.11$ & \textbf{0.3} & $3.07 \pm 0.06$ & 12.62 & 20.55 & 0.44 & $73.51 \pm 0.12$ &  \textbf{0.3} & 12.58 & 19.84 & 0.45\\
 & $73.06 \pm 0.11$ & \textbf{0.34} & $3.05 \pm 0.06$ & 11.76 & 17.97 & 0.02 & $73.06 \pm 0.13$ & \textbf{0.34} & 11.65 & 16.64 & 0.02\\
 & $72.93 \pm 0.16$ & $0.352 \pm 0.011$ & 3.06 $\pm 0.06$ & 11.22 & 14.21 & 0.10 & $72.85 \pm 0.24$ & $0.361 \pm 0.019$ & 10.48 & 10.44 &  0.18\\
 \hline

\end{tabular}
}
\caption{Best-fit values with 1 $\sigma$ uncertainties for \textit{Pantheon} and \textit{Pantheon+}, for all cosmological cases and $\cal L$. Fixed cosmological parameters are in bold. $H_0$ is in $\mathrm{km} \, \mathrm{s}^{-1} \, \mathrm{Mpc}^{-1}$.}
\label{tab:bestfit}
\end{table*}

\begin{table}
\centering
\begin{tabular}{cccccc}
\hline
\multicolumn{4}{c}{$\cal L$$_{logistic}$ vs $\cal L$$_{Gauss}$}\\
\hline
\textit{Pantheon} & Case & $\zeta_{\mathcal{L},  \Omega_{M}}$ & $\zeta_{\mathcal{L},H_0}$\\
\hline
 & $H_0=70$ & 0.4 & -  \\
 & $H_0=73.04$& 3.1 & - \\
 & $\Omega_M=0.3$ & -  & 0 \\
 & $\Omega_M=0.34$ & - & 0.2 \\
 & $\Omega_M$, $H_0$ free & 0.6 & 0.07 \\
\hline
\multicolumn{4}{c}{$\cal L$$_{Student}$ vs $\cal L$$_{Gauss}$}\\
\hline
\textit{Pantheon +} & Case & $\zeta_{\mathcal{L},\Omega_M}$ & $\zeta_{\mathcal{L},H_0}$\\
\hline
 & $H_0=70$ & 3.7 & - \\
 & $H_0=73.04$ & 0.3 & -  \\
 & $\Omega_M=0.3$ & - & 0.06  \\
 & $\Omega_M=0.34$ & - & 0\\
 & $\Omega_M$, $H_0$ free & 0.4 & 0.3\\
\hline
\end{tabular}
\caption{Computed values of $\zeta_{\mathcal{L},\Omega_M}$ and $\zeta_{\mathcal{L},H_0}$, as described in Section \ref{fit}. The first column reports the sample studied, the second one the cosmological case considered, and the other ones the values of $\zeta_{\mathcal{L},\Omega_M}$ and $\zeta_{\mathcal{L},H_0}$. $H_0$ is in units of $\mathrm{km} \, \mathrm{s}^{-1} \, \mathrm{Mpc}^{-1}$. We use the notation "-" to indicate the cases in which the z-score cannot be computed since the cosmological parameter is fixed and not fitted in the cosmological analysis.}
\label{tab:zscore:likelihood}
\end{table}

\begin{table}
\centering
\begin{tabular}{cccccc}
\hline
\multicolumn{4}{c}{$\cal L$$_{logistic}$}&\multicolumn{2}{c}{$\cal L$$_{Gauss}$}\\
\hline
\textit{Pantheon} & Case & $\zeta_{\Omega_M}$ & $\zeta_{H_0}$ & $\zeta_{\Omega_M}$ & $\zeta_{H_0}$ \\
\hline
 & $H_0=70$ & 0.2 & - & 0.2 & - \\
 & $H_0=73.04$& 0 & - & 0.1 & - \\
 & $\Omega_M=0.3$ & - & 0.1 & - & 0.3 \\
 & $\Omega_M=0.34$ & - & 0.1 & - & 0.2\\
 & $\Omega_M$, $H_0$ free & 0.1 & 0.07 & 0.06 & 0\\
\hline
\multicolumn{4}{c}{$\cal L$$_{Student}$}&\multicolumn{2}{c}{$\cal L$$_{Gauss}$}\\
\hline
\textit{Pantheon +} & Case & $\zeta_{\Omega_M}$ & $\zeta_{H_0}$ & $\zeta_{\Omega_M}$ & $\zeta_{H_0}$\\
\hline
 & $H_0=70$ & 0.1 & - & 0.1 & -\\
 & $H_0=73.04$ & 0.1 & - & 0.07 & - \\
 & $\Omega_M=0.3$ & - & 0 & - & 0.06 \\
 & $\Omega_M=0.34$ & - & 0.06 & - & 0\\
 & $\Omega_M$, $H_0$ free & 0.2 & 0.04 & 0.04 & 0.03\\
\hline
\end{tabular}
\caption{Computed values of $\zeta_{\Omega_M}$ and $\zeta_{H_0}$, as described in Section \ref{fit}, for both the new likelihoods and the Gaussian likelihood. The first column reports the sample studied, the second one the cosmological case considered, and the other ones the values of $\zeta_{\Omega_M}$ and $\zeta_{H_0}$. $H_0$ is in units of $\mathrm{km} \, \mathrm{s}^{-1} \, \mathrm{Mpc}^{-1}$. We use the notation "-" to indicate the cases in which the z-score cannot be computed since the cosmological parameter is fixed and not fitted in the cosmological analysis.}
\label{tab:zscore}
\end{table}

\section{Conclusions}
\label{conclusions}

We have investigated the Gaussianity assumption on the distance moduli $\mu$ of \textit{Pantheon} \citep{scolnic2018} and \textit{Pantheon +} \citep{pantheon+} of SNe Ia, including both statistical and systematic uncertainties. We show that this assumption, usually implicitly assumed, is not fulfilled via the Anderson-Darling and Shapiro-Wilk normality tests, and the skewness, kurtosis, and ``skewness+kurtosis" tests.
We have computed the distribution of the normalized, via the $C$ matrix, difference between the $\mu$ observed and the $\mu$ expected under the assumption of an underlying cosmological model. We have assumed a flat $\Lambda$CDM model with $\Omega_M = 0.3$ and $H_0 = 70  \, \mathrm{km} \, \mathrm{s}^{-1} \, \mathrm{Mpc}^{-1}$, but we have checked our results against this assumption testing also other specific values. We have also proved that our results do not change if we compute the theoretical $\mu$ with the general Eq. \ref{dmlcdm} in place of the Eq. \ref{dmlcdm_corr} from \citet{2019ApJ...875..145K}, which is more suitable for low redshifts. 
We have used several statistical methods to verify if $\Delta \mu _{norm}$ is normally distributed. 
Independent and complementary approaches are crucial to guarantee reliable results, leveraging on the advantages and minimizing the limits of each method.  Both SNe Ia samples fail all of these tests revealing a non-Gaussian $\Delta \mu _{norm}$ distribution. Consequently, the traditional practise of constraining cosmological parameters with SNe Ia by maximizing the Gaussian distance moduli likelihood is not appropriate, as it is not grounded on a statistical base.
We have then unveiled the hidden underlying distributions of $\Delta \mu _{norm}$, by using two different built-in functions in Wolfram Mathematica and Python, which employ several statistical tests (see Sect. \ref{fit}) to compare among numerous distributions. We notice that the best-fit distributions are the Student's t and the logistic for the \textit{Pantheon} and \textit{Pantheon +}, respectively, among all the distributions we tried. 
The logistic and the Student's t distributions are the ones that best fit \textit{Pantheon} and \textit{Pantheon +}, respectively (see Fig. \ref{fig:hist}).
This is clearly visualized in Fig. \ref{fig:hist}, where the corresponding best-fit curves (in orange) better reproduce the $\Delta \mu$ histograms of the two SNe Ia samples compared to the Gaussian best-fit curve (in green).
This result is also visible from Table \ref{tab:distributions}, in which the first five best-fitting distributions for both \textit{Pantheon} and \textit{Pantheon +} are shown along with the corresponding evaluated statistical tests. Indeed, the logistic and Student's t distributions, respectively for \textit{Pantheon} and \textit{Pantheon +}, are the favored ones according to the values of the logarithm of the likelihood, BIC, AIC, and p-values of Pearson $\chi^2$ and Cramer Von Mises tests.
Following these results, we have fitted the flat $\Lambda$CDM model using both $\cal L$$_{Gaussian}$ and the logistic ($\cal L$$_{logistic}$) and Student's t ($\cal L$$_{Student}$) likelihoods, for the \textit{Pantheon} and the \textit{Pantheon +} samples, respectively. 

Here, we summarize the main points.
The $\cal L$$_{logistic}$ and $\cal L$$_{Student}$, used to fit respectively the \textit{Pantheon} and \textit{Pantheon +} samples, with a flat $\Lambda$CDM model with both $\Omega_M$ and $H_0$ free parameters, significantly reduce the uncertainties on these parameters.
$\cal L$$_{logistic}$ reduces the uncertainties on $\Omega_M$ and $H_0$ by $ 43 \%$ (from 0.021 to 0.012) and $41 \%$ (from 0.34 to 0.20), respectively, while $\cal L$$_{Student}$ reduces the error on $\Omega_M$ and $H_0$ by $42 \%$ (from 0.019 to 0.011) and $33 \%$ (from 0.24 to 0.16), respectively. 
Our results steadily demonstrate that these new distributions perform better on the data compared to the Gaussian.
In addition, we have demonstrated that choosing the actual distribution for each SNe Ia sample is key to substantially improving the constraints on cosmological parameters.

In conclusion, we show
that $\Delta \mu _{norm}$ of \textit{Pantheon} and \textit{Pantheon +} does not verify the Gaussianity assumption. This result has a significant repercussion on the SNe Ia cosmology, the approach implicitly assumed in SNe Ia cosmology, and future SNe Ia collections for which the Gaussianity assumption will have to be carefully inspected.
In an era in which the cosmological debate is focused on tensions in the cosmological parameters, the reduction of $ \sim 40\%$ in $\Omega_M$ and $H_0$ represents a breakthrough to constrain cosmological parameters more precisely, thus shedding light on the effective extent of these tensions.

\section*{Acknowledgements}
This study uses data from \url{https://github.com/dscolnic/Pantheon} and \url{https://github.com/PantheonPlusSH0ES}.
GB acknowledges Scuola Superiore Meridionale, for supporting her visit at NAOJ, Division of Science. GB thanks the Division of Science for being hosted on campus. MGD acknowledges NAOJ. We thank M. Ghodsi Yengejeh for the work on the right panel of Fig. \ref{fig:hist}.

\section*{Authors contributions}

Conceptualization: MB, MGD, GB, SN

Data curation: GB

Formal Analysis: MGD, GB

Funding acquisition: MGD, MB 

Investigation: GB, MGD

Methodology: MGD

Software: GB, MGD

Supervision: MGD

Writing – original draft: GB, MGD

Writing – review \& editing: MGD, GB, MB, SC, SN

\section*{Conflicts of interest}
The authors declare no conflict of interest.

\section*{Data Availability}

This study uses data from \url{https://github.com/dscolnic/Pantheon} and \url{https://github.com/PantheonPlusSH0ES}.

\appendix

\section{Tests with mock samples}
\label{appendix}

\begin{figure*}
\centering
 \includegraphics[width=0.49\textwidth]{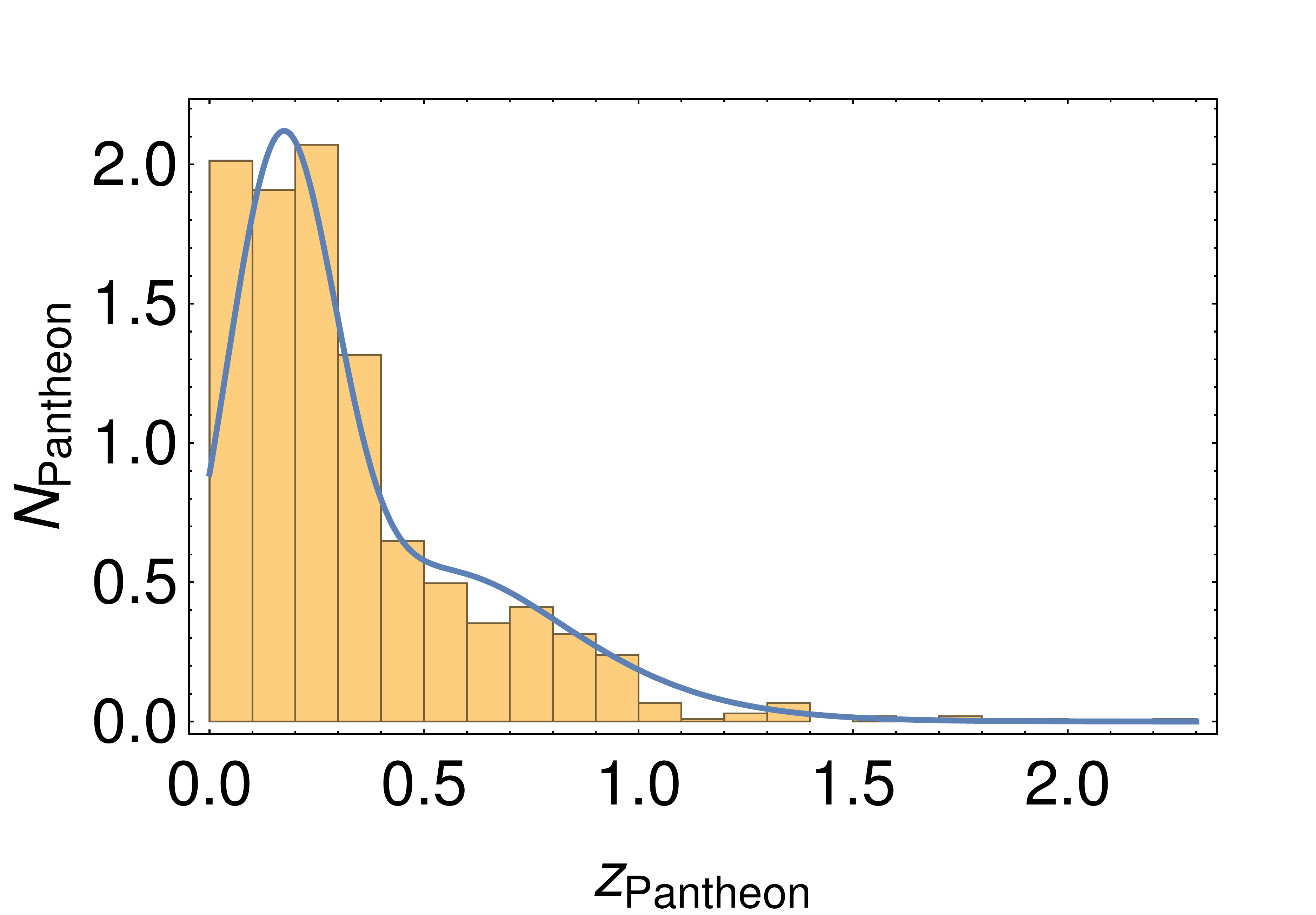}
    \includegraphics[width=0.49\textwidth]{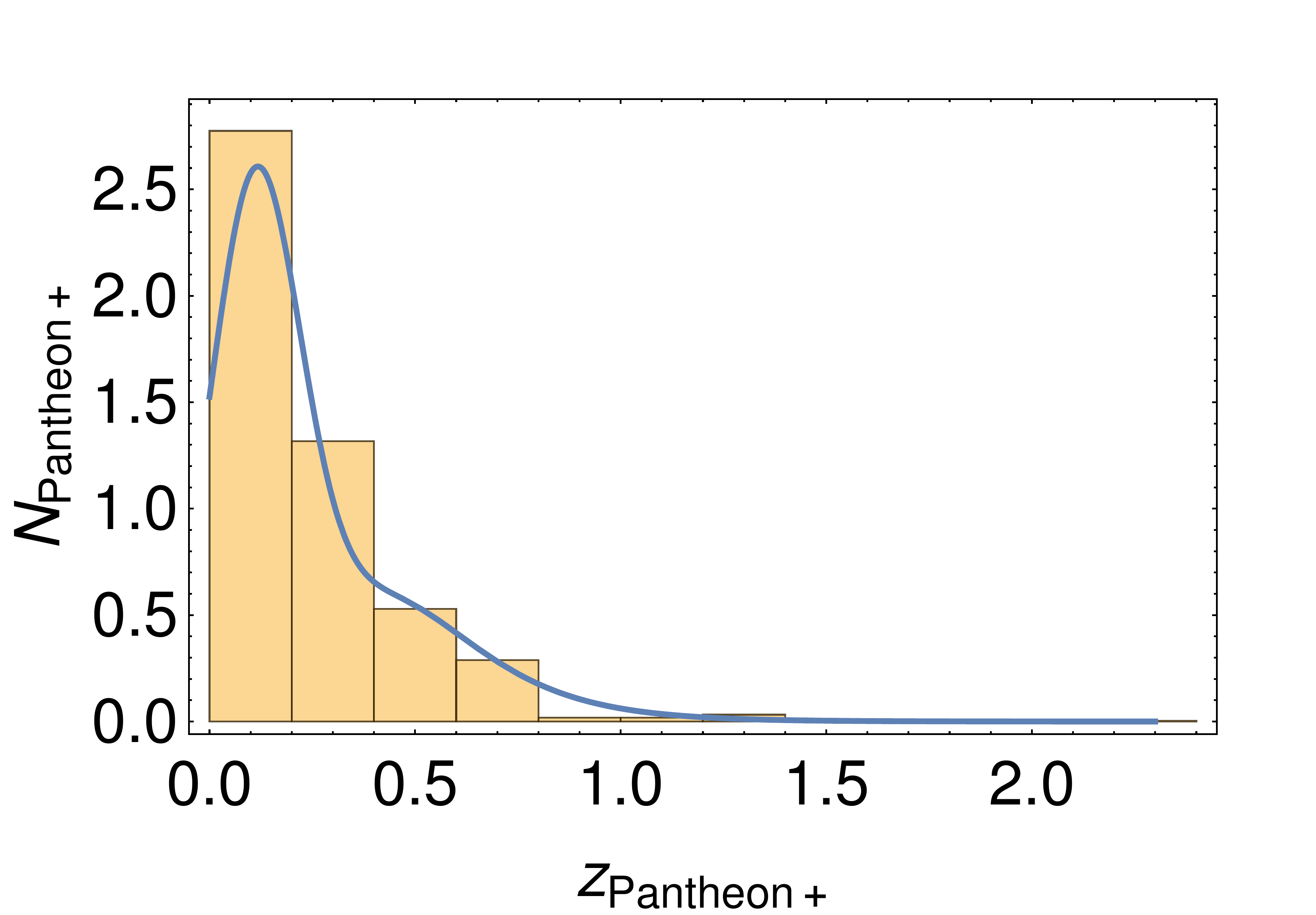}
    \includegraphics[width=0.49\textwidth]{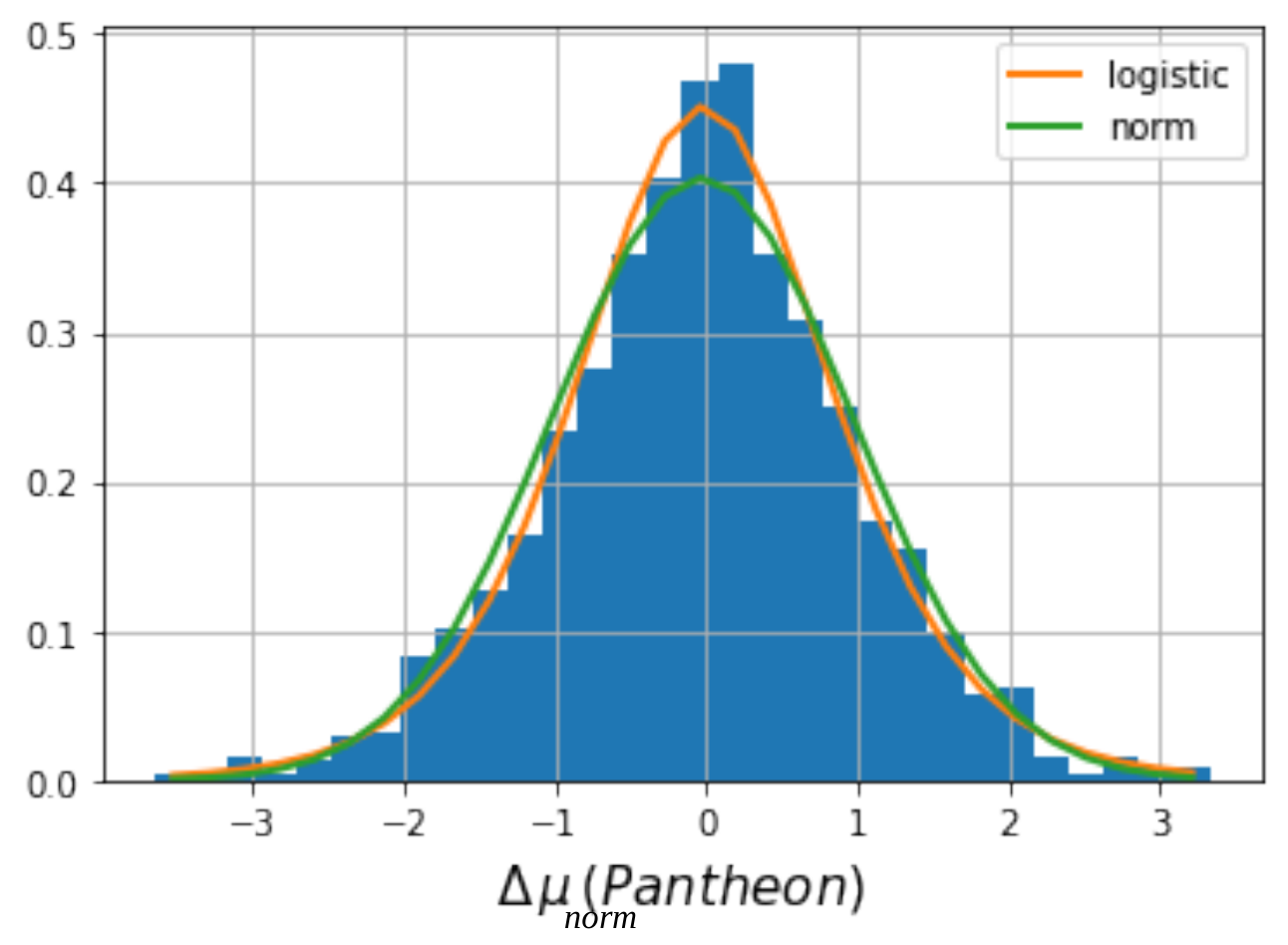}
    \includegraphics[width=0.49\textwidth]{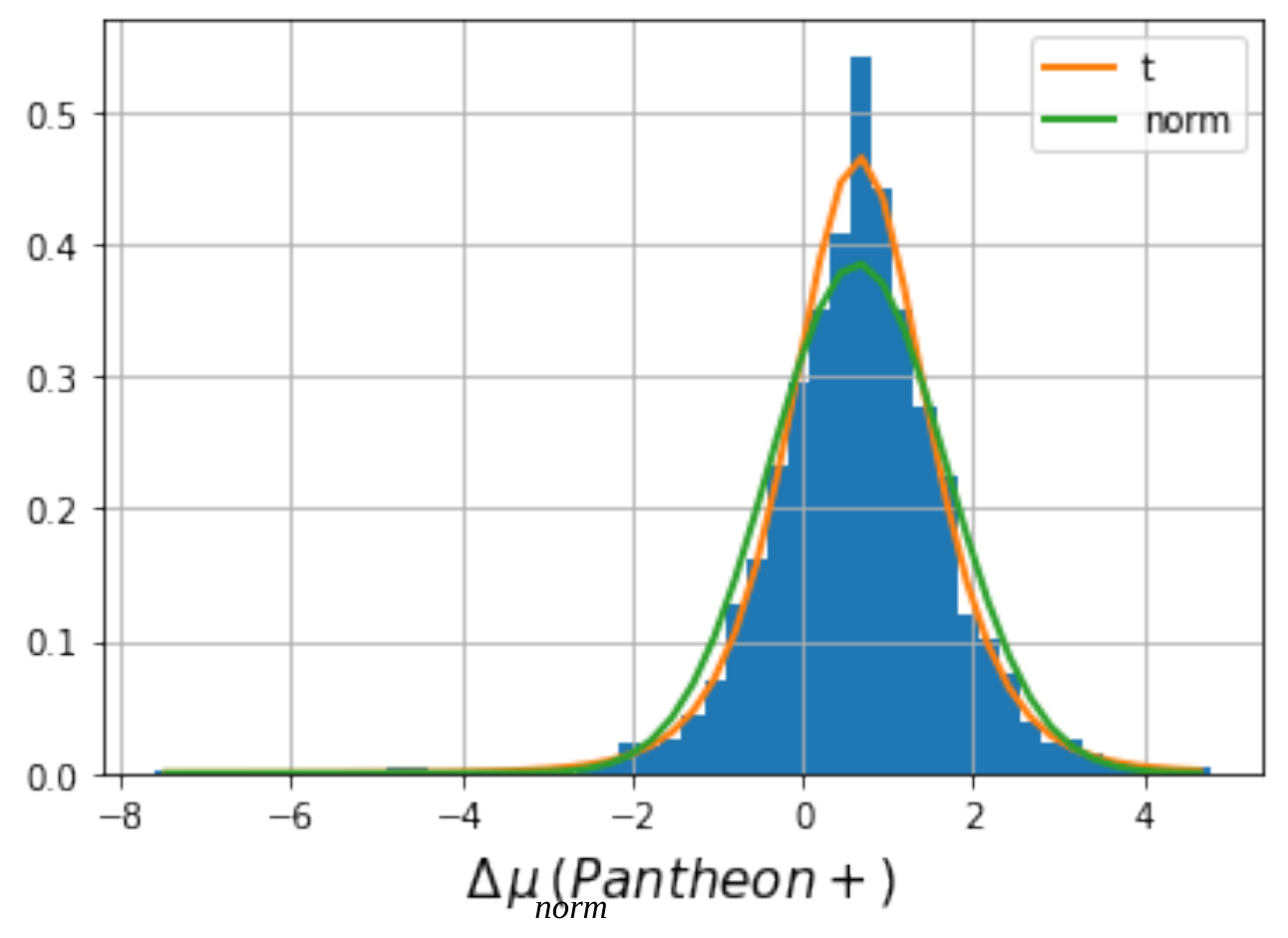}
    \caption{Upper panels: Redshift distributions for mock samples built from the 1048 \textit{Pantheon} SNe Ia  (left panel) and the 1701 \textit{Pantheon +} SNe Ia (right panel). The blue curves are the corresponding best-fit mixture distributions. Lower panels: Normalized $\Delta \mu _{norm}$ histogram for the mock samples of the 1048 SNe Ia in \textit{Pantheon} (left panel) and the 1701 SNe Ia in \textit{Pantheon +} (right panel) obtained assuming a flat $\Lambda$CDM model with $\Omega_M=0.3$ and $H_0=70 \, \mathrm{km} \, \mathrm{s}^{-1} \, \mathrm{Mpc}^{-1}$. The green curve is the best-fit Gaussian distribution, while the orange curves are the best-fit logistic (left panel) and Student's t (right panel) distributions.}
    \label{fig:histmock}
\end{figure*}

To further test the reliability of our results we have generated mock data for the redshift, the distance moduli, $\mu$, assuming a given cosmology, and the uncertainties on the distance moduli for both Pantheon (1048 mock data) and Pantheon+ (1701 mock data) samples by extracting random values from their actual distributions.
We here show that the observed redshift distribution for the Pantheon sample is a Mixture Gaussian distribution between a Normal distribution with mean=0.2 and standard deviation $\sigma=0.13$, and a $\Gamma$ distribution with the shape parameter $6.3$ and a scale of $0.1$, see the upper left panel of Fig. \ref{fig:histmock}. For the Pantheon+, we still have a Gaussian Mixture distribution between a Normal with mean=$0.12$ and $\sigma=0.11$ and a log-Normal with a mean of $-0.66$ and $\sigma=0.38$, see the upper right panel of Fig. \ref{fig:histmock}.

We have then checked that the distribution of the mock sample for $\Delta \mu _{norm}$ for the Pantheon sample obtained from the mock redshift distribution resembling the observed sample is still a logistic distribution and for the Pantheon+ sample is still a Student's T distribution. The lower left and right panels of Fig. \ref{fig:histmock} show the best-fit distributions of $\Delta \mu _{norm}$ superimposed in orange to the differential distributions for the Pantheon and Pantheon+ samples, respectively. As a comparison for both panels we have also shown the corresponding Gaussian fit in green in both panels.  

\begin{figure*}
\centering
\begin{tabular}{ccc}
  \includegraphics[width=55mm]{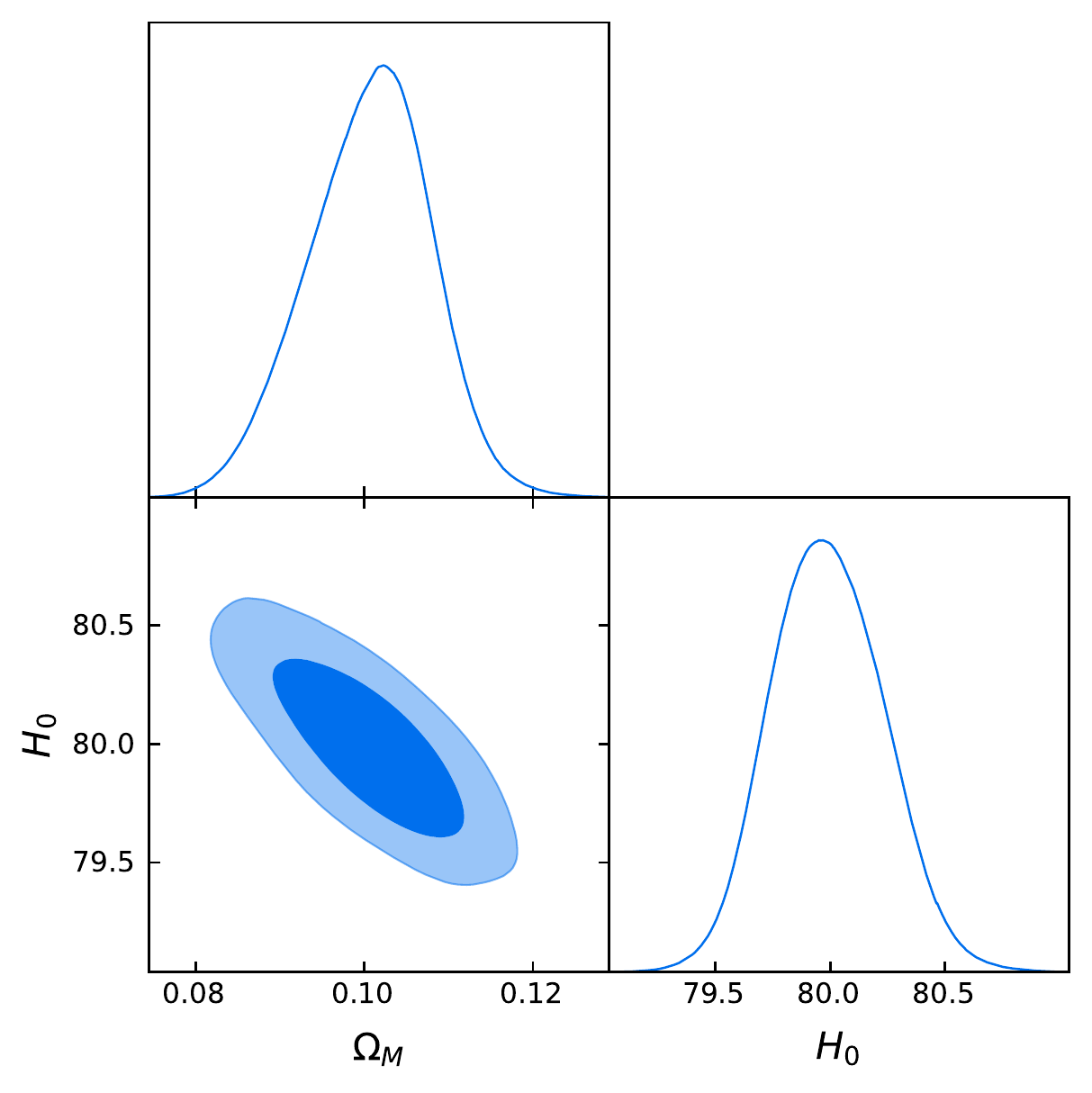}  &    \includegraphics[width=55mm]{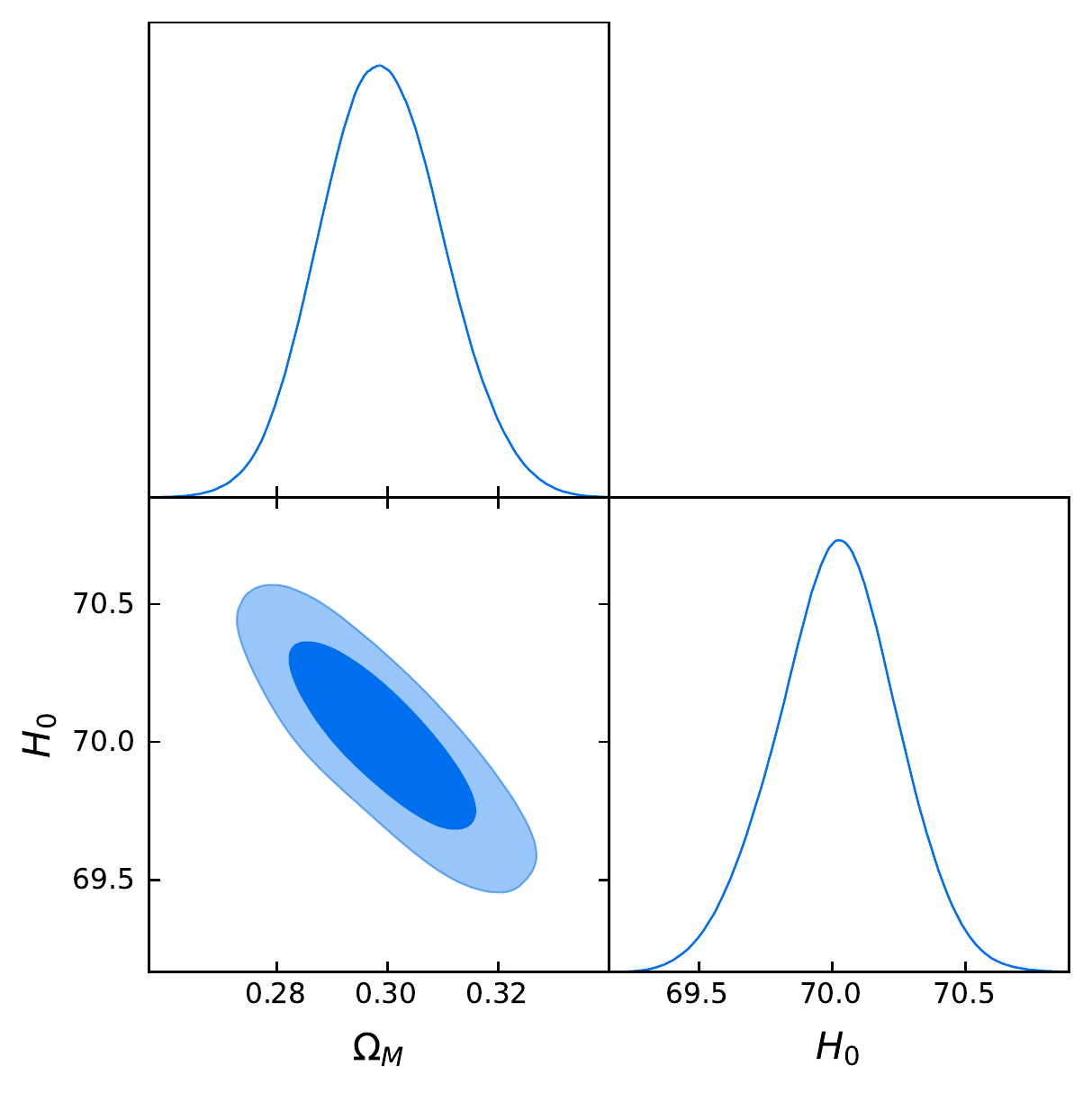}   &    \includegraphics[width=55mm]{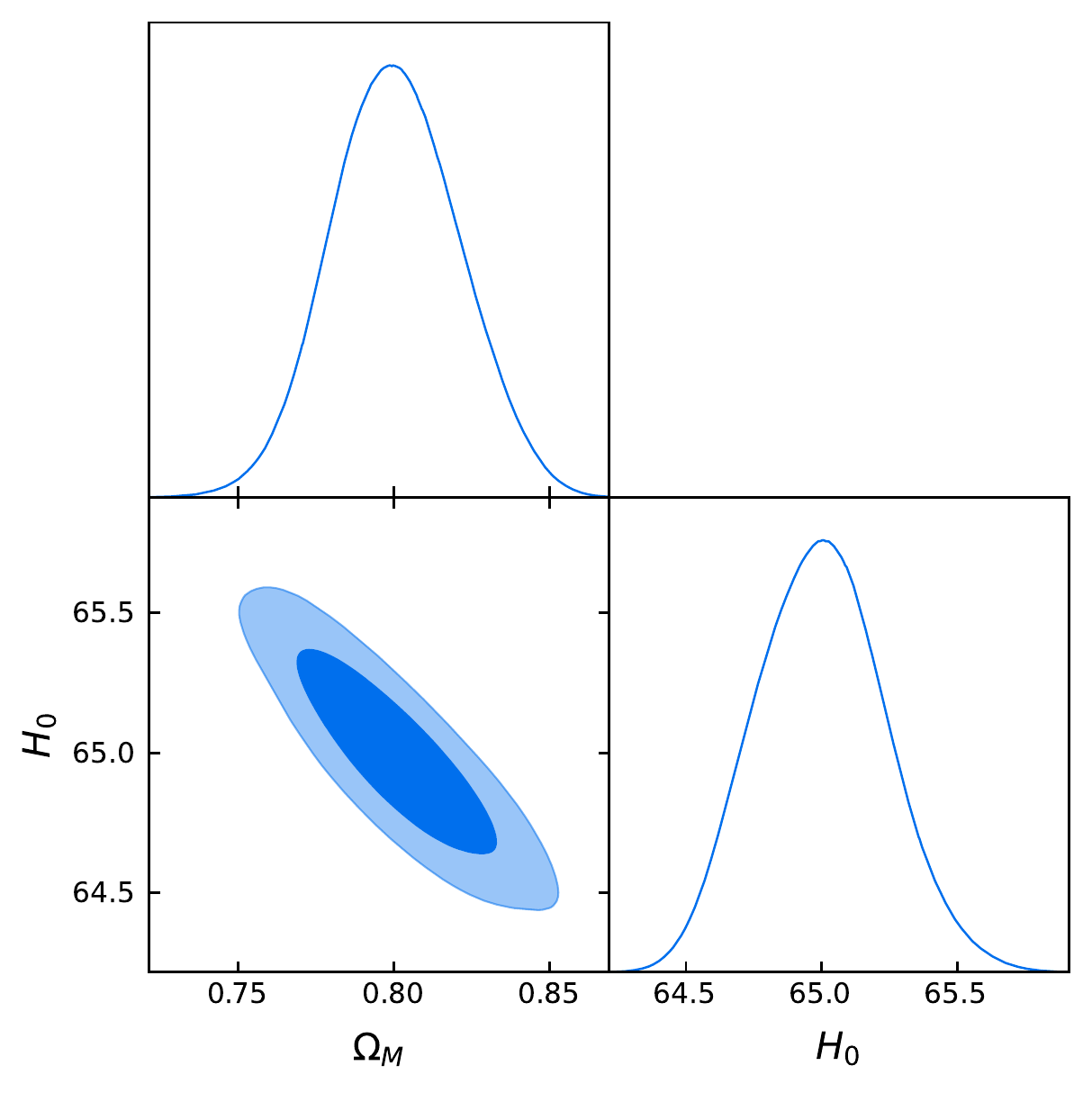}\\
  (a) $\cal L$$_{Gauss}$, $\Omega_M=0.1$, $H_0=80$  &  (b) $\cal L$$_{Gauss}$, $\Omega_M=0.3$, $H_0=70$ & (c) $\cal L$$_{Gauss}$,  $\Omega_M=0.8$, $H_0=65$
\\[6pt]
 \includegraphics[width=55mm]{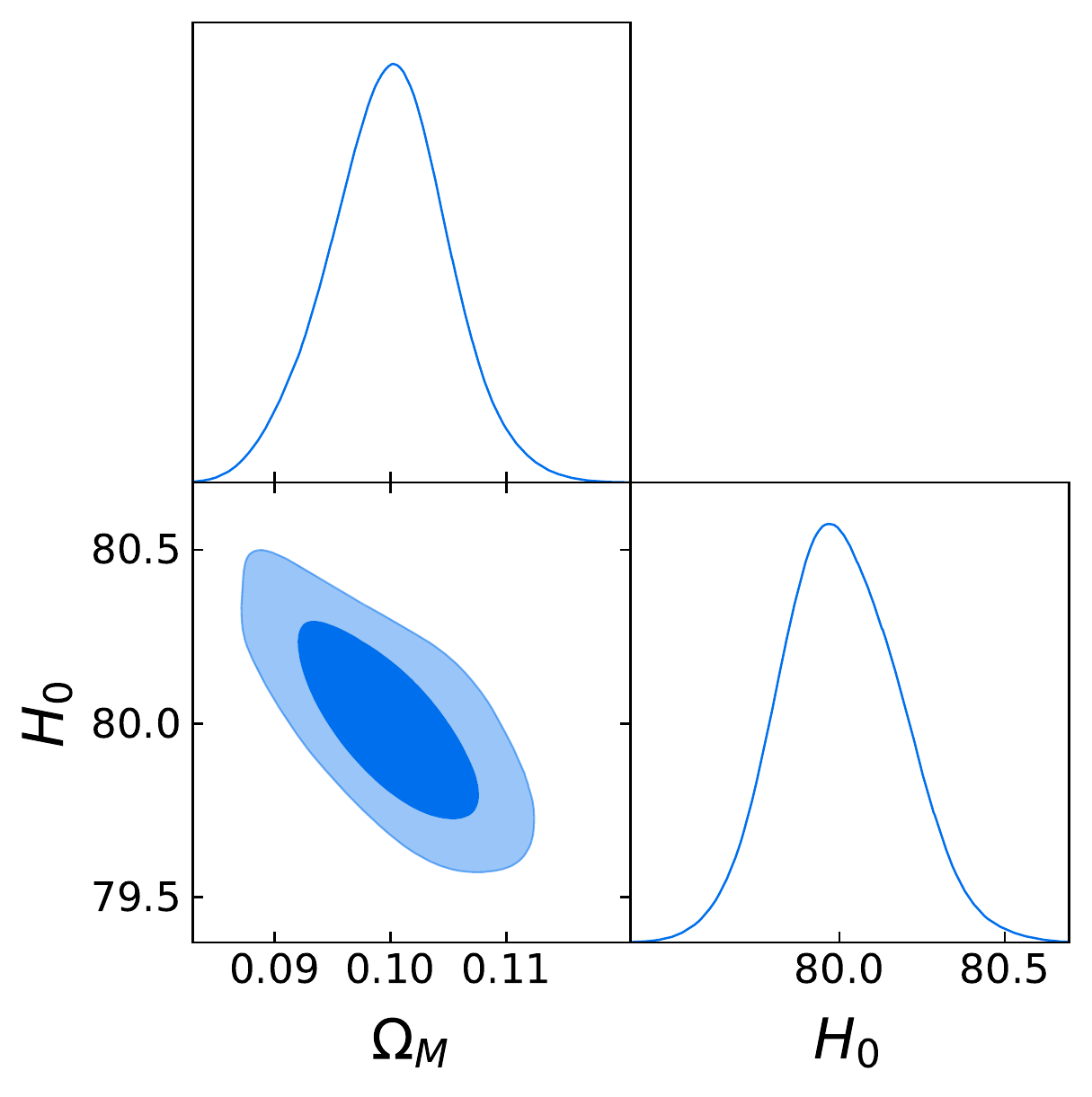} &   \includegraphics[width=55mm]{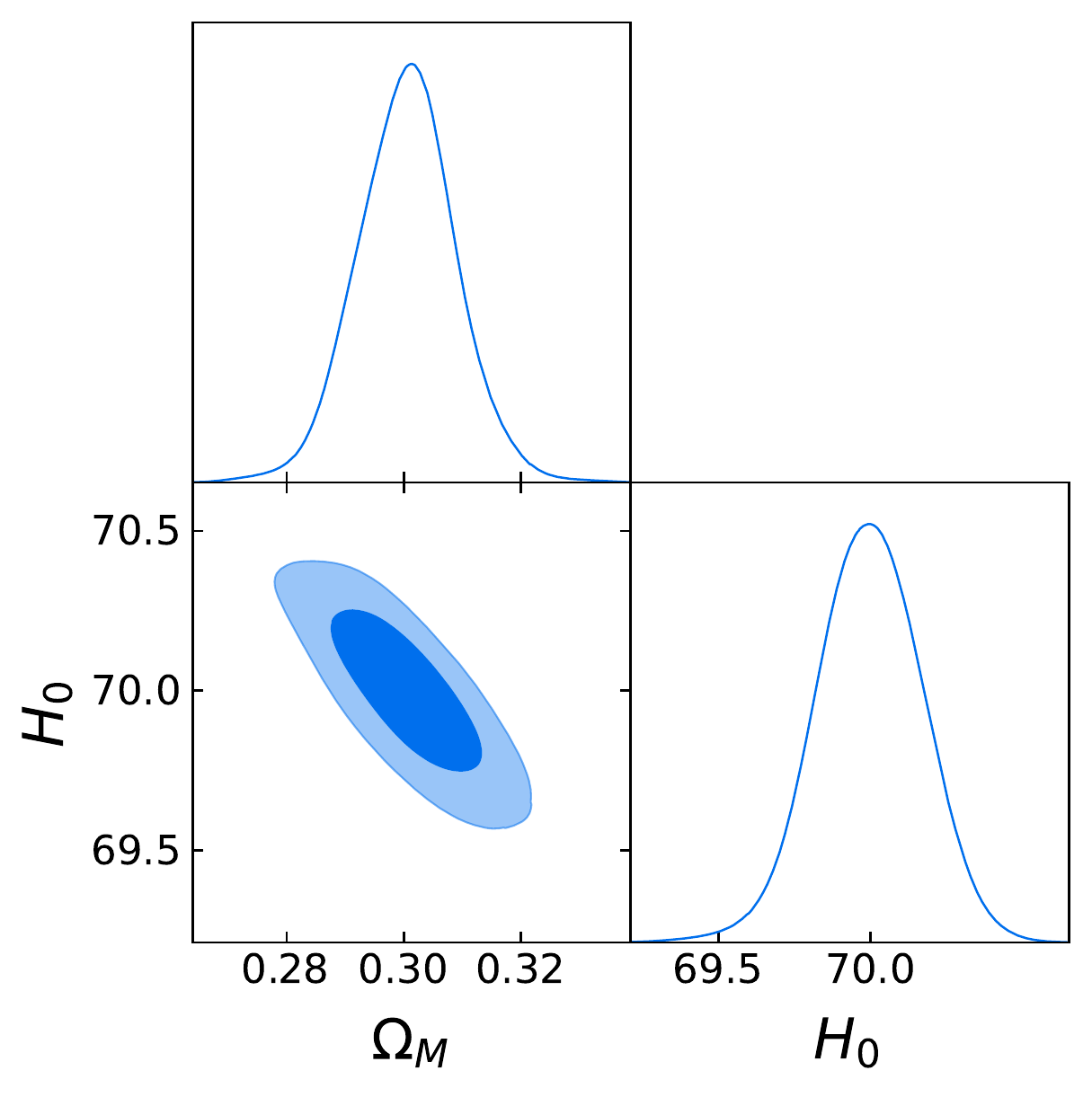} &   \includegraphics[width=55mm]{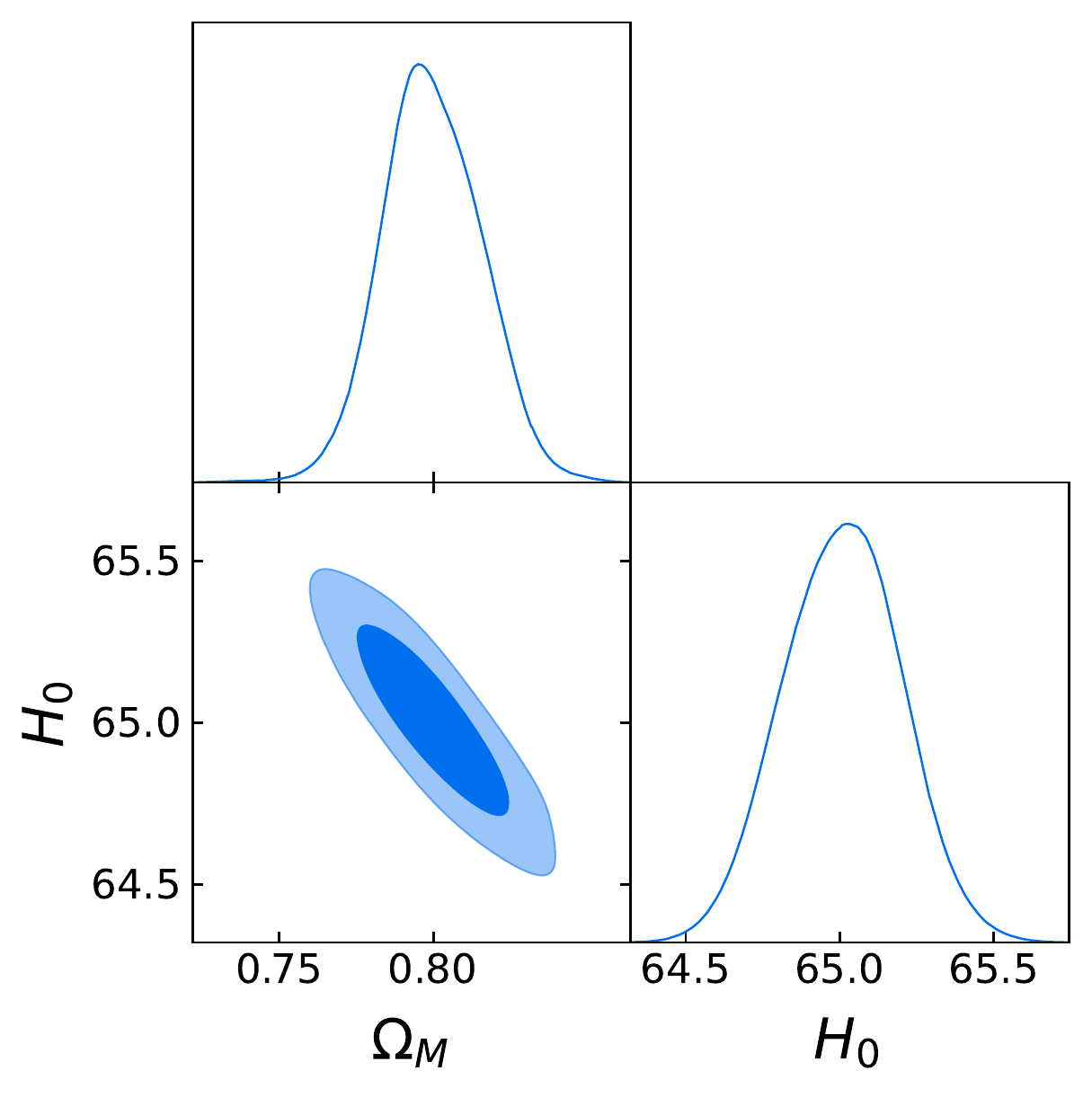} \\
  (d) $\cal L$$_{Logistic}$, $\Omega_M=0.1$, $H_0=80$ & (e)  $\cal L$$_{Logistic}$, $\Omega_M=0.3$, $H_0=70$ & (f) $\cal L$$_{Logistic}$, $\Omega_M=0.8$, $H_0=65$\\[6pt]
\end{tabular}
\caption{Fit of the flat $\Lambda$CDM model with $\Omega_M$ and $H_0$ free parameters for the mock sample for the Pantheon sample. $H_0$ is expressed in units of $\mathrm{km} \, \mathrm{s}^{-1} \, \mathrm{Mpc}^{-1}$.
}
\label{fig:histmockPantheon}
\end{figure*}

\begin{figure*}
\centering
\begin{tabular}{ccc}
  \includegraphics[width=55mm]{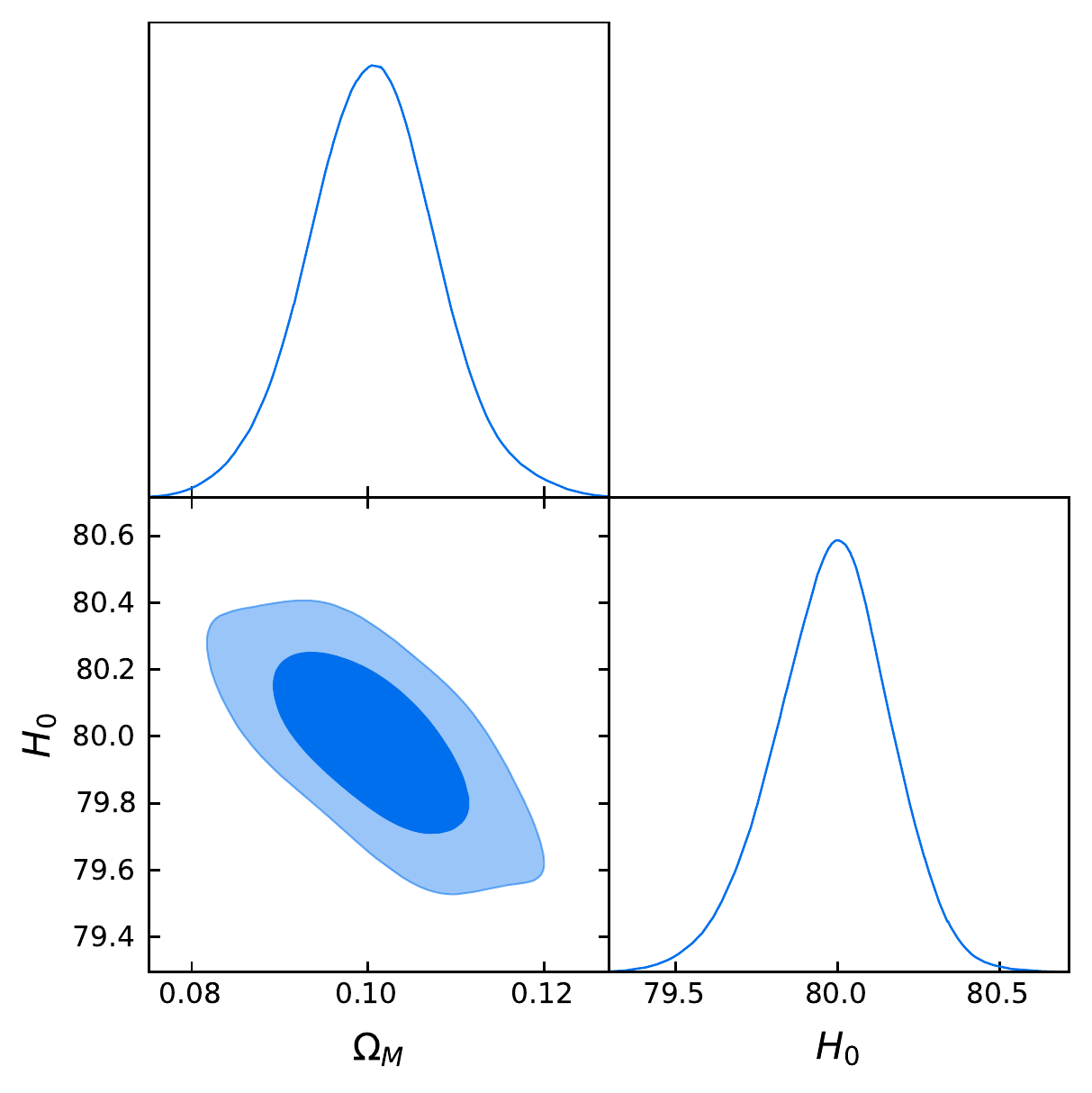}  &    \includegraphics[width=55mm]{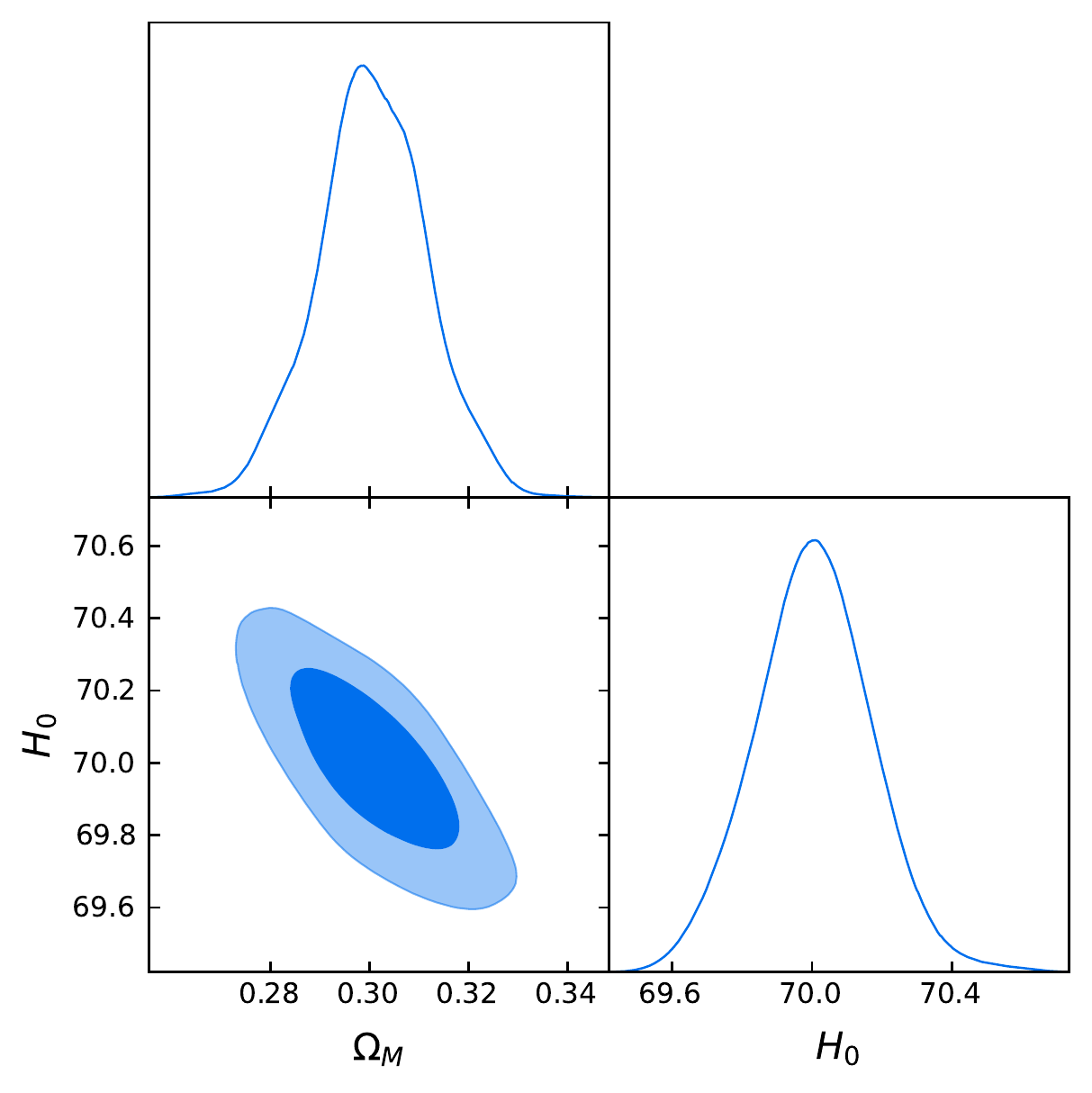}   &    \includegraphics[width=55mm]{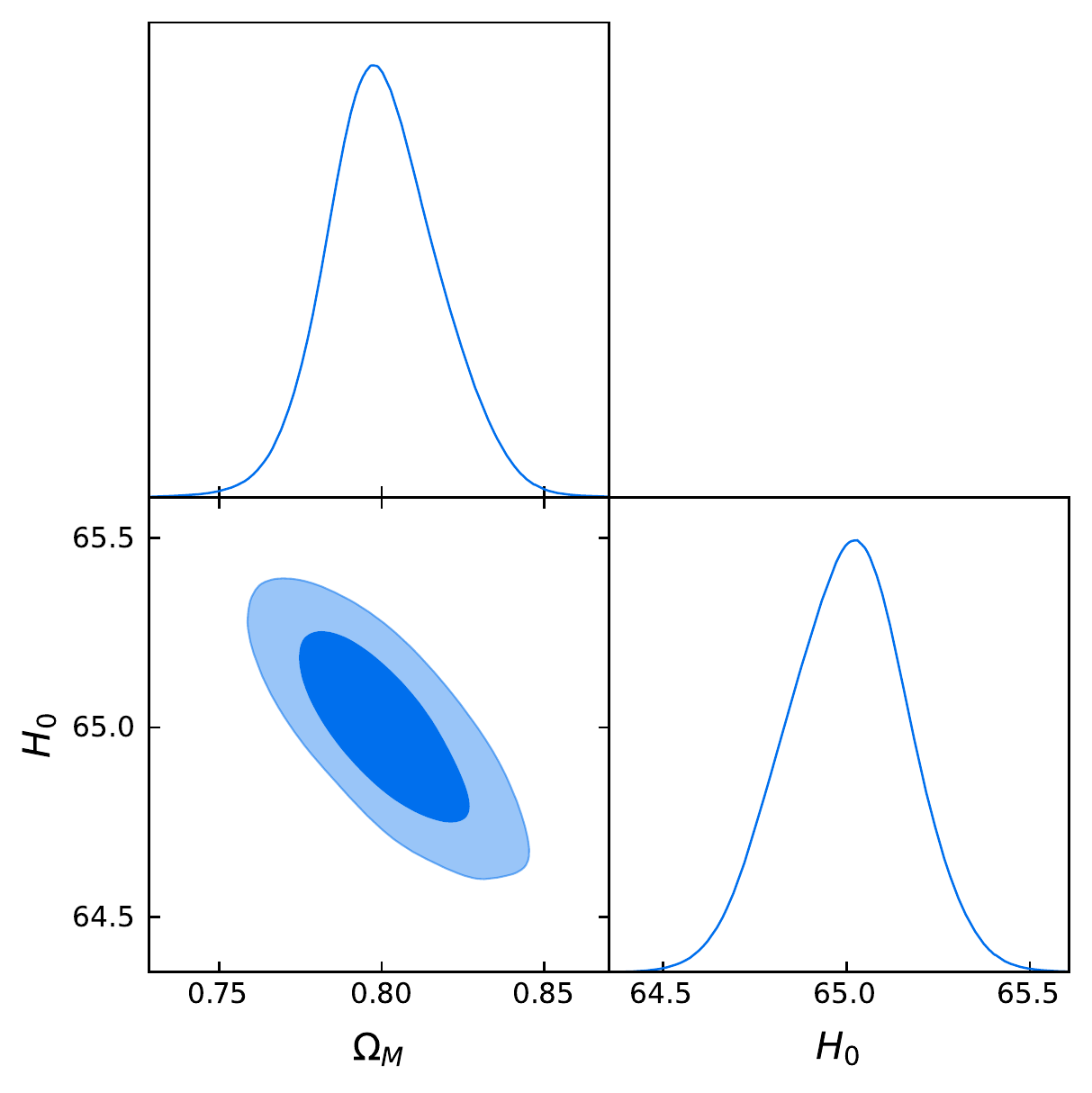}\\
(a) $\cal L$$_{Gauss}$, $\Omega_M=0.1$, $H_0=80$  &  (b) $\cal L$$_{Gauss}$, $\Omega_M=0.3$, $H_0=70$ & (c) $\cal L$$_{Gauss}$,  $\Omega_M=0.8$, $H_0=65$  \\[6pt]
 \includegraphics[width=55mm]{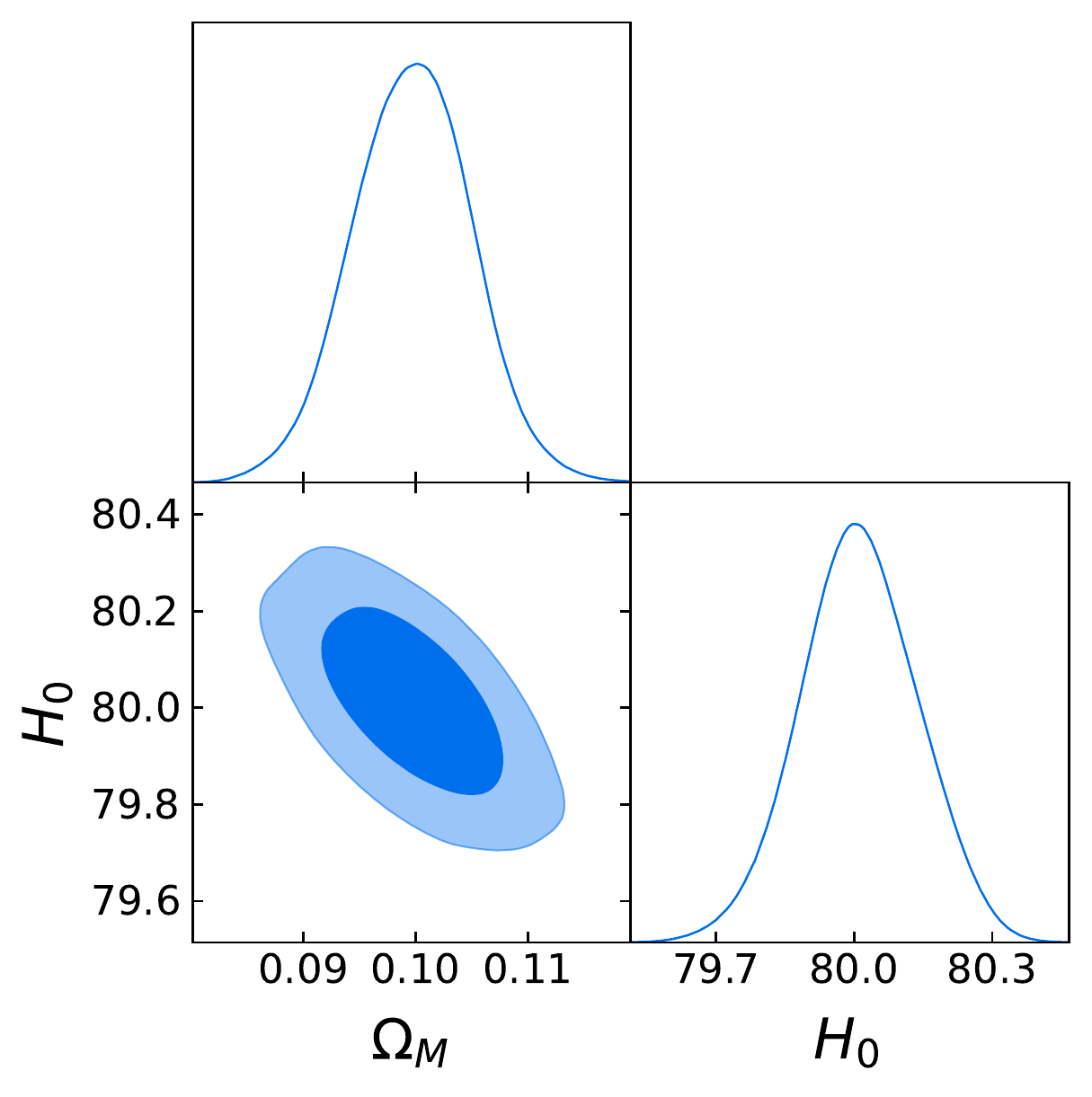} &   \includegraphics[width=55mm]{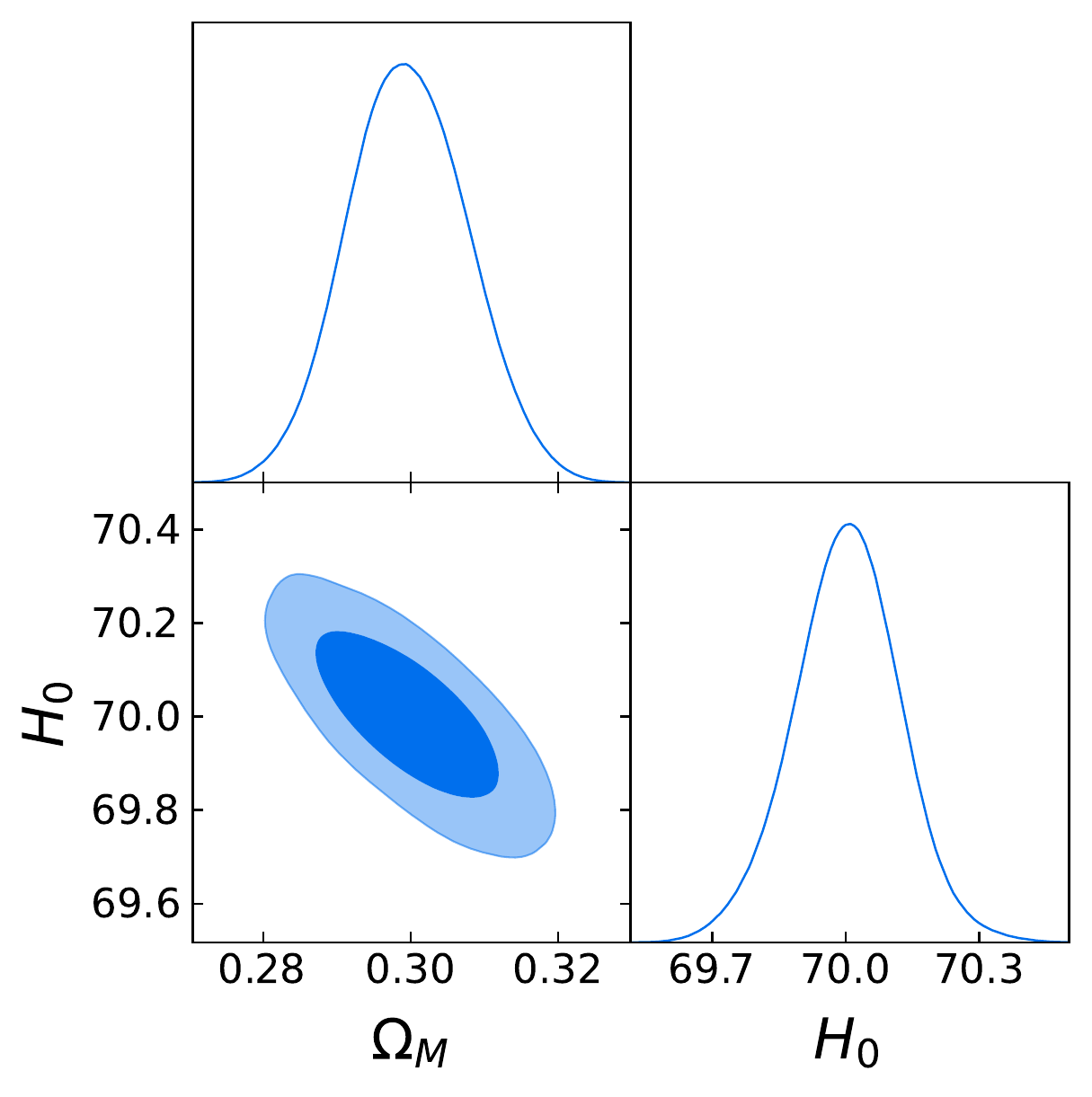} &  \includegraphics[width=55mm]{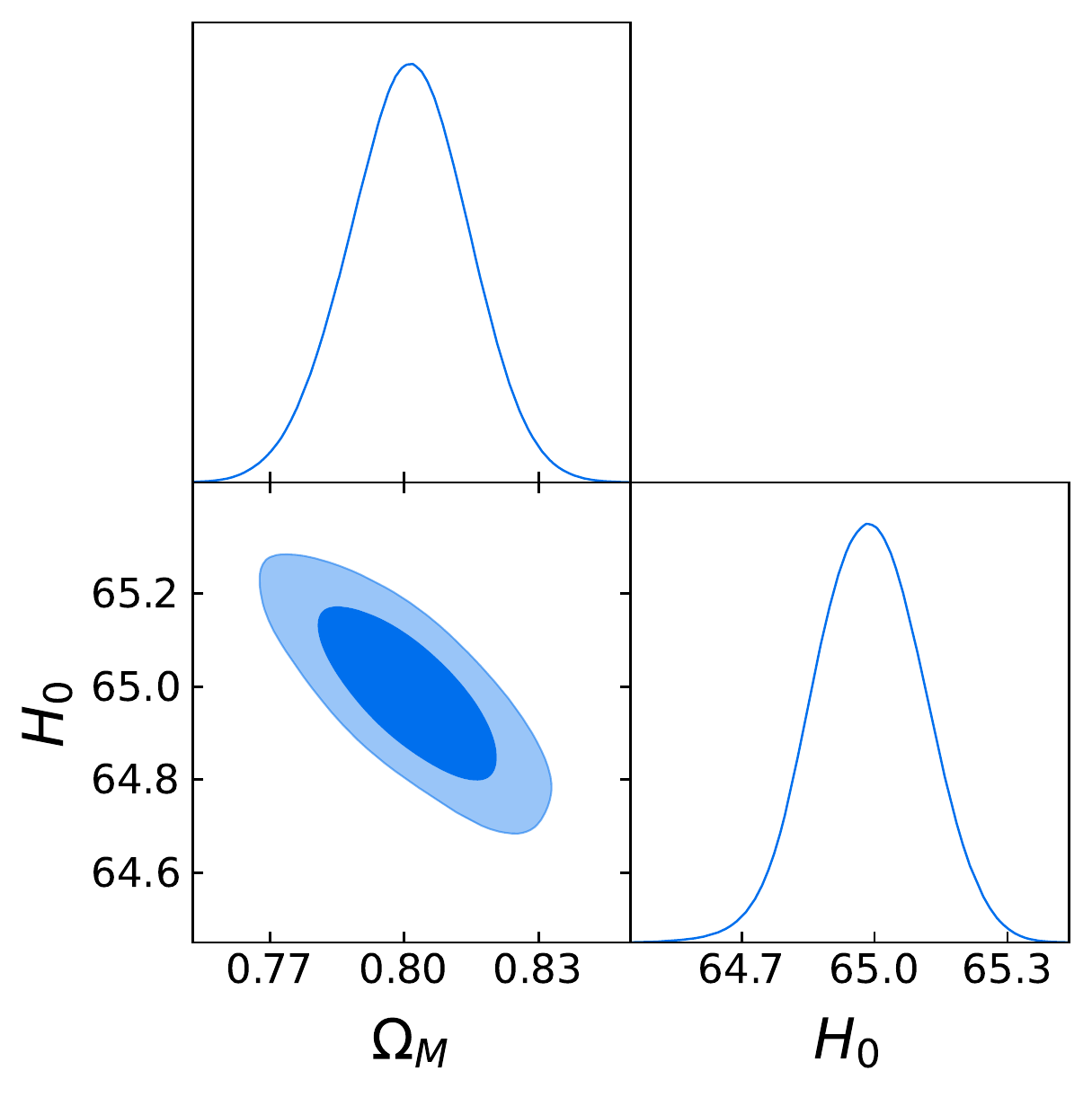} \\
(d) $\cal L$$_{Student}$, $\Omega_M=0.1$, $H_0=80$ & (e)  $\cal L$$_{Student}$, $\Omega_M=0.3$, $H_0=70$ & (f) $\cal L$$_{Student}$, $\Omega_M=0.8$, $H_0=65$ \\[6pt]
\end{tabular}
\caption{Fit of the flat $\Lambda$CDM model with $\Omega_M$ and $H_0$ free parameters for the mock sample for the Pantheon+ sample. $H_0$ is expressed in units of $\mathrm{km} \, \mathrm{s}^{-1} \, \mathrm{Mpc}^{-1}$.
}
\label{fig:histmock_Pantheon+}
\end{figure*}

We have then computed the $\mu$ with the theoretical formula in Eq. \eqref{dmlcdm_corr} assuming a given cosmological model in the flat $\Lambda$CDM model. Specifically, we have tested three cases: $\Omega_M=0.3$ and $H_0=70 \, \mathrm{km} \, \mathrm{s}^{-1} \, \mathrm{Mpc}^{-1}$, $\Omega_M=0.8$ and $H_0=65 \, \mathrm{km} \, \mathrm{s}^{-1} \, \mathrm{Mpc}^{-1}$, and $\Omega_M=0.1$ and $H_0=80 \, \mathrm{km} \, \mathrm{s}^{-1} \, \mathrm{Mpc}^{-1}$.
We have fitted the mock data leaving together $\Omega_M$ and $H_0$ free to vary both with the Gaussian and the Logistic likelihood, see the upper and lower panels of Fig. \ref{fig:histmockPantheon}, respectively, for Pantheon SNe Ia. We have fitted the Gaussian and the Student's T likelihoods for Pantheon+ SNe Ia again leaving both $\Omega_M$ and $H_0$ free to vary, see the upper and lower panels of Fig. \ref{fig:histmock_Pantheon+}, respectively. As it is shown in the figures, all results with the observed distributions are confirmed by the ones obtained with the mock samples.

\bibliographystyle{elsarticle-harv} 
\bibliography{bibliografia_4}






\end{document}